\documentclass[lettersize,journal]{IEEEtran}
\usepackage{amsmath}
\usepackage{amsfonts}
\usepackage{algorithmic}
\usepackage{algorithm}
\usepackage{array}
\usepackage{multirow}
\usepackage[caption=false,font=normalsize,labelfont=sf,textfont=sf]{subfig}
\usepackage{textcomp}
\usepackage{stfloats}
\usepackage{url}
\usepackage{verbatim}
\usepackage{graphicx}
\usepackage{cite}
\usepackage{epstopdf}
\usepackage{color}
\usepackage{threeparttable}
\usepackage{diagbox}
\usepackage{amssymb}
\usepackage{makecell}
\usepackage{hyperref}

\begin{document}

\title{Amplitude-Domain Reflection Modulation for Active RIS-Assisted Wireless Communications}
\author{Jing Zhu, \textit{Member, IEEE},  Qu Luo, \textit{Member, IEEE}, Zheng Chu, \textit{Member, IEEE}, Gaojie Chen, \textit{Senior Member, IEEE},  Pei Xiao, \textit{Senior Member, IEEE}, Lixia Xiao, \textit{Member, IEEE}, and Chaoyun Song, \textit{Senior Member, IEEE}

\thanks{
J. Zhu, Q. Luo, G. Chen, and P. Xiao are with 5G and 6G Innovation centre, Institute for Communication Systems (ICS) of University of Surrey, Guildford, GU2 7XH, UK (e-mail: \{j.zhu, q.u.luo, gaojie.chen, p.xiao\}@surrey.ac.uk).

Z. Chu is with the Department of Electrical and Electronic Engineering, University of Nottingham Ningbo China, Ningbo 315100, China, and also with the Next Generation Internet of Everything Laboratory, University of Nottingham Ningbo China, Ningbo 315100, China. (Email: andrew.chuzheng7@gmail.com).  

%P. Gao is with Pengcheng Laboratory, Shenzhen, 518055, China. (Email: gaopy@pcl.ac.cn).

Lixia Xiao is with the Research Center of 6G Mobile Communications, School of Cyber Science and Engineering, Huazhong University of Science and Technology, Wuhan 430074, China. (e-mail: lixiaxiao@hust.edu.cn).

Chaoyun Song is with the Department of Engineering, King’s College London, WC2R 2LS London, U.K. (e-mail: chaoyun.song@kcl.ac.uk).
}

}

\maketitle

\begin{abstract}
In this paper, we propose a novel active reconfigurable intelligent surface (RIS)-assisted amplitude-domain reflection modulation (ADRM) transmission scheme, termed as ARIS-ADRM. This innovative approach leverages the additional degree of freedom (DoF) provided by the amplitude domain of the active RIS to perform index modulation (IM), thereby enhancing spectral efficiency (SE) without increasing the costs associated with additional radio frequency (RF) chains.  Specifically, the ARIS-ADRM scheme transmits information bits through both the modulation symbol and the index of active RIS amplitude allocation patterns (AAPs). To evaluate the performance of the proposed ARIS-ADRM scheme, we provide an achievable rate analysis and derive a closed-form expression for the upper bound on the average bit error probability (ABEP). Furthermore, we formulate an optimization problem to construct the AAP codebook, aiming to minimize the ABEP. Simulation results demonstrate that the proposed scheme significantly improves error performance under the same SE conditions compared to its benchmarks. This improvement is due to its ability to flexibly adapt the transmission rate by fully exploiting the amplitude domain DoF provided by the active RIS. 

\end{abstract}


\section{Introduction}
\IEEEPARstart{R}{econfigurable} intelligent surfaces (RIS) has garnered substantial interest in beyond fifth-generation (B5G) communications for its potential to significantly enhance the performance of wireless networks \cite{di2020smart,wu2019intelligent,elmossallamy2020reconfigurable}. Among RIS variants, passive ones stand out for their ability to manipulate electromagnetic waves without active transmissions. This manipulation is achieved through a surface composed of numerous passive elements capable of adjusting signal phase, facilitating control over signal direction and strength. While offering cost-effective and energy-efficient solutions for signal coverage enhancement, interference mitigation, and spectral efficiency (SE) improvement, passive RIS introduces multiplicative fading, meaning the signal's path loss from the base station (BS) to the receiver through the RIS panel is affected by the product of the BS-RIS and RIS-receiver links, rather than their sum \cite{ozdogan2019intelligent}. As a result, the expected achievable gain for passive RIS-based systems will be significantly reduced. A large passive RIS with hundreds of reflecting elements (REs) can mitigate the multiplicative fading of the cascaded channel and enhance beamforming gain \cite{najafi2020physics}. However, this setup results in significant signaling overhead for channel estimation and increases computational complexity, making practical implementation challenging.

To address this issue, the concept of active RIS has been introduced in \cite{you2021wireless,long2021active,zhang2022active}. Unlike passive RIS that reflects signals without boosting them, active RISs integrate active components, such as amplifiers, into the RIS framework, allowing them to actively manipulate the signal's amplitude in addition to its phase. %This active manipulation overcomes the attenuation issues associated with passive RIS, enabling stronger and more reliable signal reflections.
The authors of \cite{you2021wireless} explored the fundamental differences between active and passive RISs, underscoring the potential of active RIS to enhance signal strength and extend communication range through amplification, a capability that passive RISs lack. The author of \cite{long2021active} further elaborated on this advantage, demonstrating that active RIS can significantly improve overall system performance by increasing received signal power and providing better reliability. In \cite{zhang2022active}, a comprehensive comparison between active and passive RIS was provided, concluding that active RISs offer superior performance in terms of signal enhancement and communication quality, making them more suitable for sixth-generation (6G) applications. Subsequently, in \cite{zhi2022active}, the performance of active RIS and passive RIS was theoretically compared under the same power budget, reinforcing the notion that active RISs are more effective in real-world scenarios. The optimization of active RIS-aided systems has been another critical area of research. The authors of \cite{ma2023optimization} proposed a novel dual-function active RIS architecture for multi-user communication systems. This architecture enables both reflection and transmission functionalities, integrating signal amplification to enhance the quality-of-service (QoS) for all users and expand coverage. 
Channel estimation of active RIS was investigated in \cite{chen2023channel}, where a least-square (LS)-based method was applied for channel estimation. 
It is worth mentioning that the aforementioned studies focused on the fully-connected  active RIS architecture, where each RE incorporates a dedicated power amplifier (PA), resulting in high power consumption. To overcome this problem, the authors of \cite{liu2021active} proposed a sub-connected active RIS architecture, where the REs are grouped into panels, sharing the same amplitude but different phase shifts. Thus, the sub-connected active RIS architecture requires fewer PAs and reduced signaling overhead compared to the fully-connected configurations.

Another promising technique is index modulation (IM), which improves spectral efficiency by encoding additional information into the indices of antennas/subcarriers/time slots/channel states \cite{zhu2023design,dang2017adaptive,wen2021joint,zhu2023index}. IM leverages the existing dimensions of communication systems to transmit extra data without requiring additional power or bandwidth. A notable example of IM is spatial modulation (SM), extensively investigated in \cite{mesleh2008spatial}. SM introduces an additional dimension for information transmission by utilizing the active antenna index, thereby improving data rates and system performance. To further optimize spectral efficiency and reliability, researchers have explored various SM derivatives. These include generalized spatial modulation (GSM) \cite{xiao2014low}, which extends SM by activating multiple antennas simultaneously; quadrature spatial modulation (QSM) \cite{li2016generalized}, which incorporates the use of both in-phase and quadrature components; enhanced spatial modulation (ESM) \cite{yang2019enhanced}, which offers improved performance through advanced signal processing techniques; and receive spatial modulation (RSM) \cite{zhu2018low}, which shifts the modulation focus to the receiver side for added benefits. Each of these variants builds upon the foundational principles of SM to address specific challenges and enhance overall system capabilities.

To explore their synergies,  several studies in the literature have focused on integrating passive RIS with IM \cite{basar2020reconfigurable,zhang2021large,zhu2022index,ma2020large,singh2022ris,guo2020reflecting,lin2020reconfigurable,lin2021reconfigurable,li2022reconfigurable}.
In particular, the authors of \cite{basar2020reconfigurable} introduced the concept of traditional IM to RIS techniques, proposing it as a new paradigm beyond traditional multiple-input multiple-output (MIMO) for 6G networks. This work laid the foundation for utilizing RIS to achieve higher SE and improved performance by dynamically altering the reflection patterns to encode information. In \cite{zhang2021large}, a novel constellation phase rotation-aided GSM was designed for large intelligent surface (LIS)-aided communication systems to achieve more robust error performance. The authors of \cite{zhu2022index} first incorporated IM into simultaneously transmitting and reflecting RIS (STAR-RIS) system, where subsurfaces are indexed to convey extra information bits. Moreover, the performance analysis of the RIS-aided IM schemes has been investigated in \cite{ma2020large,singh2022ris}. More recently, a novel IM technique, reflection modulation (RM), was proposed in \cite{guo2020reflecting}, where the phase shift patterns are indexed to convey additional information bits and all REs are activated to achieve the maximum diversity gain. In recent RM systems, passive RIS REs have been utilized with an ON/OFF keying mechanism to facilitate information transmission \cite{lin2020reconfigurable,lin2021reconfigurable,li2022reconfigurable}. This approach involves strategically activating and deactivating specific REs to modulate the reflected signals, effectively embedding data within the reflected waveforms.
The authors of \cite{yao2023superimposed} and \cite{yao2022universal} proposed a novel superimposed RIS-phase modulation (SRPM) scheme for MIMO systems, which superimposes tunable phase offsets onto RIS phases to carry additional information, demonstrating superior performance in transmission rate, bit error rate (BER), and channel capacity compared to existing RM and passive beamforming schemes.

Apart from the above passive RIS-aided IM schemes, the integration of IM with active RIS was explored in \cite{sanila2023joint}. In this study, a joint spatial and reflecting modulation (JSRM) scheme was proposed for active RIS-assisted MIMO communications, termed as ARIS-JSRM. Similar to conventional IM-aided passive RIS systems, the ARIS-JSRM scheme switches only a subset REs to the ON state to convey extra information bits. This approach, however, does not fully utilize the extra degrees of freedom provided by active RIS. To the best of our knowledge, there is a paucity of research focusing on exploring the amplitude dimension offered by active RIS to perform IM. Inspired by the passive RIS-empowered phase shift-domain RM scheme \cite{guo2020reflecting}, we propose a novel active RIS-aided amplitude-domain RM scheme to enhance the system performance.
The primary contributions of this paper can be outlined as follows.
\begin{itemize}
   \item 
   We propose a novel active RIS-enabled amplitude-domain RM transmission mechanism, called ARIS-ADRM, which fully exploits the amplitude domain diversity provided by active RIS. Specifically, in the proposed scheme, the information bits are transmitted through both the modulation symbol and the index of ARIS amplitude allocation patterns (AAP). 
   \item 
   We derive a closed-form expression for the upper bound on the average bit error probability (ABEP) and provide a comprehensive achievable rate analysis. These theoretical analyses offer valuable insights into the performance limits and capabilities of the proposed scheme.
   \item
   We formulate an optimization problem to generate the AAP codebook with the objective of minimizing error performance, which is equivalent to maximizing the minimum Euclidean distance. To address this non-convex quadratically constrained quadratic program (QCQP), we apply the successive convex approximation (SCA) method, transforming it into a series of linear convex problems for tractability.
   \item 
   We present comprehensive simulation results to confirm the benefits of the proposed scheme. The key findings are summarized as follows: 1) the accuracy of the theoretical analysis for the upper bound on the ABEP is rigorously verified; 2) the proposed ARIS-ADRM scheme is superior to its relevant benchmark schemes.
\end{itemize}

The reminder of this paper is organized as follows. Section II presents the channel model as well as the transceiver model of the proposed ARIS-ADRM scheme. Section III provides the derivation for the ABEP and achievable rate performance. The AAP codebook design is given in Section IV. This is followed by the simulation results and analysis in Section V. Finally, Section VI provides the concluding remarks of this paper.

\textit{Notations:}  $\left|  \cdot  \right|$ and ${\left\| {\bf{A}} \right\|}$ refer to the absolute value and the Frobenius norm, respectively. ${\left(  \cdot  \right)^H}$ and ${\left(  \cdot  \right)^T}$ are the Hermitian transpose and transpose, respectively.
$\rm{diag}\left(  \cdot  \right)$, $\rm{Tr}\left(  \cdot  \right)$ and ${\mathbb{E}}\left(  \cdot  \right)$ are the diagonal, trace and expectation operations, respectively. ${\Re} \left\{  \cdot  \right\}$ represents the real part of a complex variable. ${{\mathbb{C}}^{{M} \times {N}}}$ and ${{\mathbb{R}}^{{M} \times {N}}}$ denote the space of $M \times N$ complex and real matrices, respectively. $Q(x) = \frac{1}{{\sqrt {2\pi } }}\int_x^\infty  {\exp ( - {u^2}/2)} {\rm{d}}u$ is the Gaussian $Q$-function. $\angle(x)$ represents the angle response of $x$. ${{\bf{I}}_N}$ refers to an $N$-dimensional identity matrix.
$\odot$ and $\otimes$ denote the Hadamard and Kronecker products, respectively.

\begin{figure*}[ht]
\centering
\includegraphics[width=0.8\textwidth]{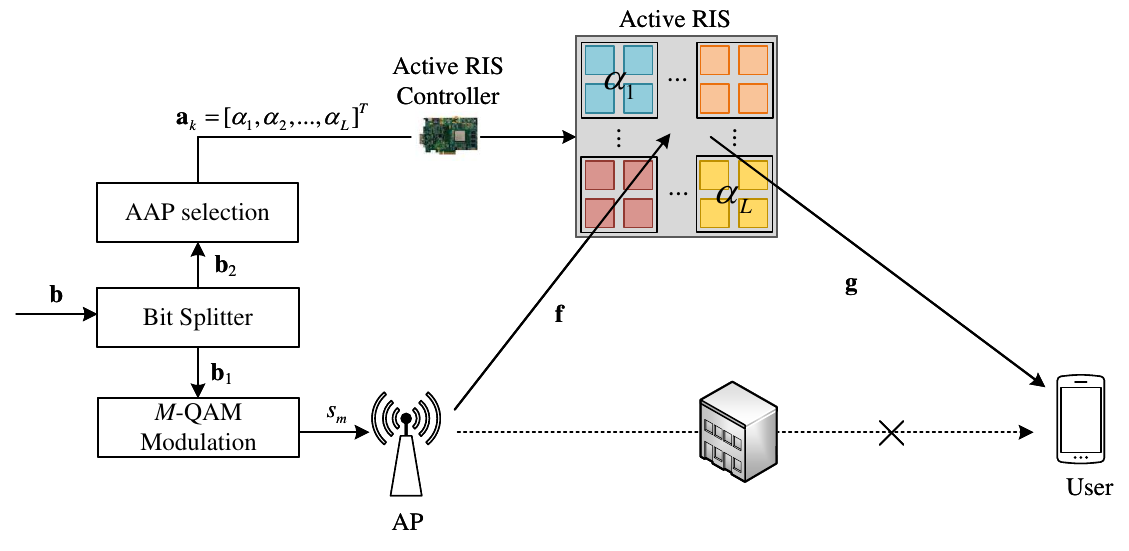}
\caption{Block diagram of the proposed ARIS-ADRM system.}\label{System_Model}
\end{figure*}

\section{System Model}
In an active RIS-assisted single-input single-output (SISO) communication system, a single-antenna Access Point (AP) communicates with a single-antenna user via an active RIS. This RIS comprises $N$ REs under the control of a central controller, enabling adjustment of both reflection phase shifts and amplitudes \cite{zhang2022active}. To efficiently implement IM, active RIS adopts the sub-connected architecture. Specifically,
the $N$ REs of the active RIS are organized into $L$ groups, where each group consists of ${\bar N} = N/L$ adjacent REs. In the $l$th group, the $i$th RE is denoted as the $(l,i)$th RE, characterized by its reflection coefficient ${\phi _{li}} = {{\alpha}_{li}}\exp (j{\theta _{li}})$. Here, ${\theta}_{li} \in [0,2\pi),l=1,2,..,L, i=1,2,..,{\bar N}$ represents the phase shift, and ${{\alpha}_{li}}>1$ signifies the amplitude due to the amplification provided by its amplifiers. 

It is assumed in this study that all REs within each group share a common amplitude $\alpha_l$ but differ in phase shift ${\theta _{li}}$ \cite{sanila2023joint}. Therefore, the reflection coefficient of the $l$th group can be alternatively expressed as ${\phi _{li}} = {{\alpha}_{l}}\exp (j{\theta _{li}})$. This configuration allows the active RIS to dynamically adjust the phase and amplitude of reflected signals, optimizing the communication channel between the AP and the user.

\subsection{Channel Model}
As shown in Fig. 1, direct links between the AP and the user are obstructed by obstacles. A more general case, which considers the presence of a direct MIMO channel between the AP and the user, will be explored in Section V. The communication channel from the AP and the active RIS is represented as ${{\bf{f}}} = {[{f_{11}},...,{f_{li}},...,{f_{L\bar N}}]^T\in {\mathbb C}^{N\times 1}}$, where $f_{li}$ refers to the channel coefficient from the AP to the $(l,i)$th active RIS RE. Similarly,
the channel between the active RIS and the user is described as ${\bf g}={[{g_{11}},...,{g_{li}},...,{g_{L\bar N}}]^T}\in {\mathbb C}^{N\times 1}$ with $g_{li}$ denoting the channel coefficient from the $(l,i)$th active RIS RE to the user. In this work, we consider a quasi-static block Rician fading channel model to describe all individual links. In particular, $f_{li}$ and $g_{li}$ are modeled as follows \cite{li2022reconfigurable}
\begin{equation}\label{H_1}
{f_{li}} = \sqrt {{\rho _1}} \left( {\sqrt {\frac{{{K_1}}}{{1 + {K_1}}}} {{\bar f}_{li}} + \sqrt {\frac{1}{{1 + {K_1}}}} {{\tilde f}_{li}}} \right)
\end{equation}
and 
\begin{equation}\label{G_1}
{g_{li}} = \sqrt {{\rho _2}} \left( {\sqrt {\frac{{{K_2}}}{{1 + {K_2}}}} {{\bar g}_{li}} + \sqrt {\frac{1}{{1 + {K_2}}}} {{\tilde g}_{li}}} \right),
\end{equation}
respectively, where the components ${{\bar f}_{li}} = \exp ( - {{j2\pi {d_1}} \mathord{\left/
 {\vphantom {{j2\pi {d_1}} \lambda }} \right.
 \kern-\nulldelimiterspace} \lambda })$ and ${{\bar g}_{li}} = \exp ( - {{j2\pi {d_2}} \mathord{\left/
 {\vphantom {{j2\pi {d_2}} \lambda }} \right.
 \kern-\nulldelimiterspace} \lambda })$ correspond to the line-of-sight (LOS) segments of the AP-to-ARIS and ARIS-to-User links, respectively. Here, $\lambda$ denotes the wavelength, $d_1$ is the distance between the AP and the active RIS, and $d_2$  denotes the distance between the active RIS and the user. Additionally, ${{\tilde f}_{li}} \sim {\mathcal{CN}}(0,1)$ and ${{\tilde g}_{li}} \sim {\mathcal{CN}}(0,1)$ are the non-line-of-sight (NLOS) components of the AP-to-ARIS and ARIS-to-user links, respectively. Moreover, $K_1$ and $K_2$ denote the Rician factors associated with the AP-to-ARIS and ARIS-to-user links, respectively, which account for the strength of the LOS component relative to the scattered components. Finally, ${\rho _1} = {\rho _r}d_1^{ - {v_1}}$ and ${\rho _2} = {\rho _r}d_2^{ - {v_2}}$ represent the large-scale path loss of the AP-to-ARIS and ARIS-to-user links, respectively. Here, $\rho_r$ is the path-loss factor at a reference distance of 1 meter (m), and $v_1$ and $v_2$ denote the path-loss exponents of the AP-to-ARIS and ARIS-to-user links, respectively, characterizing how the signal strength diminishes with distance.

\begin{comment}
{\renewcommand\arraystretch{1.5}
\begin{table}
\centering
{\caption{An example look-up table for index bits with $L=4$.}}
\begin{tabular}{|c|c|c||c|c|c|}\hline
${\bf{b}}_2$ & Index & AAP $\cal A$ & ${\bf{b}}_2$ & Index & AAP $\cal A$  \\ \hline
0000 & 1 &  $[p_1,p_2,p_3,p_4]$   & 1000 & 9 & $[p_2,p_3,p_1,p_4]$     \\ \hline
0001 & 2 & $[p_1,p_2,p_4,p_3]$   & 1001 & 10 & $[p_2,p_3,p_4,p_1]$     \\ \hline
0010 & 3 & $[p_1,p_3,p_2,p_4]$   & 1010 & 11 & $[p_2,p_4,p_1,p_3]$     \\ \hline
0011 & 4 & $[p_1,p_3,p_4,p_2]$   & 1011 & 12 & $[p_2,p_4,p_3,p_1]$     \\ \hline
0100 & 5 & $[p_1,p_4,p_2,p_3]$   & 1100 & 13 & $[p_3,p_1,p_2,p_4]$     \\ \hline
0101 & 6 & $[p_1,p_4,p_3,p_2]$   & 1101 & 14 & $[p_3,p_1,p_4,p_2]$     \\ \hline
0110 & 7 & $[p_2,p_1,p_3,p_4]$   & 1110 & 15 & $[p_3,p_2,p_1,p_4]$     \\ \hline
0111 & 8 & $[p_2,p_1,p_4,p_3]$   & 1111 & 16 & $[p_3,p_2,p_4,p_1]$     \\ \hline
\end{tabular}
\end{table}}
\end{comment}

\subsection{Signal Model}
In the proposed scheme, the concept of IM is employed on the active RIS side, where index bits are conveyed by the AAP of each active RIS group. The amplitude coefficient for each active RIS group is dynamically configured according to the selected AAP, a process overseen by the active RIS controller. This controller receives instructions from the AP through a feedforward link, enabling precise adjustment of the RIS configuration based on the transmitted index bits information. Notably, a key feature of this scheme is the simultaneous activation of all active RIS REs to maximize transmitted power. This operational strategy contrasts with the approach described in \cite{sanila2023joint}, where the activation pattern and power allocation may differ significantly. This distinction highlights the unique capability of the proposed system to optimize power usage and enhance signal transmission efficiency through coordinated RIS control mechanisms.

\begin{comment}
As depicted in Fig. \ref{System_Model}, the incoming data stream is divided into two sub-vectors: ${\bf b}_1$ containing $\log_2{M}$ bits and ${\bf b}_2$ containing ${{\log }_2}(A)$ bits, where $M$ and $A$ are the modulation order and the AAP codebook order, respectively. The ${\bf b}_1$ vector encodes symbol bits that are modulated into an $M$-QAM symbol $s_m$ transmitted from the AP, where $s_m \in {\cal S}, m=1,2...,M$ and $\cal S$ are the $M$-QAM constellation set. The index bits in ${\bf b}_2$ are used to select a unique AAP vector ${\bf{a}}_k=[\alpha_1,...,\alpha_L]^T \in {\mathbb{C}}^{L \times 1} $ from a pre-defined AAP codebook, i.e., ${\cal A}=[{\bf{a}}_1,...,{\bf{a}}_k,...,{\bf{a}}_A]\in {\mathbb R}^{L \times A},k=1,2,...,A$. Each AAP vector ${\bf{a}}_k$ determines the specific amplitude coefficients for all the $L$ ARIS groups.
\end{comment}

As depicted in Fig. \ref{System_Model}, the incoming data stream is divided into two sub-vectors: ${\bf b}_1$ containing $\log_2{M}$ bits and ${\bf b}_2$ containing ${{\log }_2}(A)$ bits, where $M$ and $A$ are the modulation order and the AAP codebook order, respectively. The detailed mapping rule is given as follows
\begin{itemize}
    \item  
    The ${\bf b}_1$ vector encodes symbol bits that are modulated into an $M$-QAM symbol $s_m$ transmitted from the AP, where $s_m \in {\cal S}, m=1,2...,M$ and $\cal S$ are the $M$-QAM constellation set.
    \item 
    The index bits in ${\bf b}_2$ are used to select a unique AAP vector ${\bf{a}}_k=[\alpha_1,...,\alpha_L]^T \in {\mathbb{C}}^{L \times 1} $ from a pre-defined AAP codebook, i.e., ${\cal A}=[{\bf{a}}_1,...,{\bf{a}}_k,...,{\bf{a}}_A]\in {\mathbb R}^{L \times A},k=1,2,...,A$. Each AAP vector ${\bf{a}}_k$ determines the specific amplitude coefficients for all the $L$ ARIS groups.
\end{itemize}
Consequently, the transmission rate of the proposed ARIS-ADRM scheme is computed as
\begin{equation}\label{Data_rate}
R=    \log_2{M}+ \log_2{A}.
\end{equation}

{\renewcommand\arraystretch{1.5}
\begin{table}
\centering
{\caption{An example look-up table for $A=4$ and $M=2$.}}
\begin{tabular}{|c|c|c|c|}\hline
\thead{Bits at AP \\ ${\bf{b}}_1$} & \thead{Bits at RIS \\ ${\bf{b}}_2$}  & \thead{Transmitted \\ signal} & \thead{Activated \\ AAP } \\ \hline
0 & 00 &  $s_1$   & ${\bf a}_1$     \\ \hline
0 & 01 & $s_1$   & ${\bf a}_2$      \\ \hline
0 & 10 & $s_1$   & ${\bf a}_3$      \\ \hline
0 & 11 & $s_1$   & ${\bf a}_4$     \\ \hline
1 & 00 & $s_2$   & ${\bf a}_1$      \\ \hline
1 & 01 & $s_2$   & ${\bf a}_2$     \\ \hline
1 & 10 & $s_2$   & ${\bf a}_3$     \\ \hline
1 & 11 & $s_2$   & ${\bf a}_4$     \\ \hline
\end{tabular}
\end{table}}

%To illustrate the mapping rule more intuitively, Table I provides a simple example with system parameters $M=2$ and $A=4$. In this case, ${{\log}_2}M=1$ and $\log_2{A}=2$, indicating ${\bf b}_1$ contains 1 bit and ${\bf b}_2$ contains 2 bits. For instance,  if ${\bf b}_1=[0]$ and ${\bf b}_2=[01]$, the transmitted signal from the AP is $s_1$, and the selected AAP for the active RIS groups is ${\bf a}_2$. In practical implementation, optimizing the amplitude coefficients for each active RIS group under a specified power budget is crucial, which will be explored further in Section IV. This optimization ensures efficient utilization of resources and enhances system performance in the ARIS-ADRM framework.

To illustrate the mapping rule more intuitively, Table I provides a simple example with system parameters $M=2$, $A=4$ and $L=2$. In this case, ${{\log}_2}M=1$ and $\log_2{A}=2$, indicating ${\bf b}_1$ contains 1 bit and ${\bf b}_2$ contains 2 bits. For instance,  if ${\bf b}_1=[0]$ and ${\bf b}_2=[01]$, the transmitted signal from the AP is $s_1$, and the selected AAP for the active RIS groups is ${\bf a}_2$, which assigns specific amplitude coefficients $[\alpha_1,\alpha_2,...,\alpha_L]^T$ to the $L$ active RIS groups.

Suppose the AAP codebook is pre-defined as
\begin{equation}
{\cal A} = \left[ {\begin{array}{*{20}{l}}
1&{0.8}&{0.6}&{0.4}\\
{0.4}&{0.6}&{0.8}&1
\end{array}} \right].
\end{equation}
Here, each column represents a unique AAP vector. If ${\bf b}_2=[01]$, the selected vector is ${\bf{a}}_2=[0.8,0.6]^T$. This configuration specifies the amplitude coefficients for the active RIS groups, where the first group has an amplitude coefficient of 0.8 and the second group has 0.6. In practical implementation, optimizing these amplitude coefficients under a specified power budget is crucial. Such optimization ensures efficient utilization of resources and enhances system performance in the ARIS-ADRM framework. This aspect will be explored further in Section IV, focusing on the trade-off between power allocation and system reliability.

%Table I presents an example of the mapping between the index bits in ${\bf b}_2$ and the AAP $\cal A$ with $L=4$, where the length of ${\bf b}_2$ is $\left\lfloor {{{\log }_2}(4!)} \right\rfloor=4$. Here, the possible amplitudes for all the $4$ active RIS groups are assumed to be $\{p_1,p_2,p_3,p_4\}$. %{\footnote{For a given power budget, the possible amplitude for each active RIS group can be optimized, which will be discussed in the next section.}}. 
%When ${\bf b}_2=[0011]$, the corresponding AAP ${\cal A}=[p_1,p_3,p_4,p_2]$. Thus, in this case, the amplitude for each active RIS group is set to ${\alpha}_1=p_1, {\alpha}_2=p_3, {\alpha}_3=p_4$ and ${\alpha}_4=p_2$.

At the user end, the received signal $y$ is expressed as
\begin{equation}\label{received_signal}
\begin{aligned}
y = &\sqrt {{P_{\rm{AP}}}} \left( {\sum\limits_{l = 1}^L {\sum\limits_{i = 1}^{\bar N} {{\bf{a}}_k(l){f_{li}}{g_{li}}\exp (j{\theta _{li}})} } } \right)s_m \\
&\quad + \sum\limits_{l = 1}^L {\sum\limits_{i = 1}^{\bar N} {{{\bf{a}}_k(l)}{g_{li}}\exp (j{\theta _{li}})} } {n_r} + n,
\end{aligned}
\end{equation}
where $P_{\rm{AP}}$ is the transmit power of the AP, $n_r \sim {\cal{CN}}(0,\sigma_r^2)$ is the thermal noise introduced by active RIS components, and $n \sim {{\cal{CN}}(0,\sigma_0^2)}$ denotes the thermal noise at the user. \eqref{received_signal} indicates  that the information delivered to the user comprises two components: the direct signal $s_m$ from the AP and the indirectly embedded information through AAP ${\bf a}_k$ from the active RIS. This dual-component transmission strategy enhances flexibility and efficiency in data delivery to the user.

To obtain the maximum instantaneous signal-to-noise ratio (SNR), it is optimal to set the phase shift of the $(l,i)$th RE as
${\theta _{li}} =  - \angle ({f_{li}}{g_{li}})$ \cite{basar2020reconfigurable}.
Conversely, the channel coefficients in both the AP-to-ARIS and ARIS-to-user links (i.e., \eqref{H_1} and \eqref{G_1}) can be rewritten in a polar form as ${f_{li}} = \left| {{f_{li}}} \right|\exp (j{\phi _{li}})$ and ${g_{li}} = \left| {{g_{li}}} \right|\exp (j{\varphi _{li}})$, respectively. Thus, the phase shift of the $(l,i)$th RE can be calculated by 
\begin{equation}\label{phase_shift}
{\theta _{li}} =   -({\phi _{li}}+{\varphi _{li}}).
\end{equation}
By substituting \eqref{phase_shift} to \eqref{received_signal}, the received signal $y$ can be further written as
\begin{equation}\label{Received_signal_1}
\begin{aligned}
y &= \sqrt {{P_{{\rm{AP}}}}} \left( {\sum\limits_{l = 1}^L {{\bf{a}}_k(l){h_l}} } \right)s_m + w\\
& = \sqrt {{P_{{\rm{AP}}}}} {{\bf{h}}^T}{\bf{a}}_ks_m + w , 
\end{aligned}
\end{equation}
where ${\bf h}=[h_1,...,h_l,...,h_L]^T$ with $h_l=\sum\nolimits_{i = 1}^{\bar N} {\left| {{f_{li}}} \right|\left| {{g_{li}}} \right|}$, and $w = \sum\nolimits_{l = 1}^L {\sum\nolimits_{i = 1}^{\bar N} {{\bf{a}}_k(l)\left| {{g_{li}}} \right|{e^{ - j{\phi _{li}}}}{n_r}} }  + n$. It is important to note that as $N$ increases, applying the central limit theorem (CLT) approximates the equivalent noise $w$ to a complex Gaussian random variable, denoted as $w \sim {\cal{CN}}(0,\sigma _w^2)$, where
\begin{equation}\label{noise}
\sigma _w^2 = N{\rho _2}\sigma _r^2{{\left( {\sum\nolimits_{l = 1}^L {{{({{\bf{a}}_k}(l))}^2}} } \right)} \mathord{\left/
 {\vphantom {{\left( {\sum\nolimits_{l = 1}^L {{{({{\bf{a}}_k}(l))}^2}} } \right)} L}} \right.
 \kern-\nulldelimiterspace} L} + \sigma _0^2.
\end{equation}
The detailed derivation can be found in Appendix A{\footnote{Although the active reflection elements introduce additional noise, the overall noise remains a zero-mean Gaussian distribution, ensuring it does not impact the validity of the ABEP theoretical derivation.}}.

\begin{comment}
According to the central limit theorem (CLT), when $N \gg 1$,  $h_l \sim {\cal N}(\mu,\sigma^2)$, where
\begin{equation}\label{mean_1}
\mu  = \bar N{\mathbb E}\{ \left| {{f_{li}}} \right|\} {\mathbb E}\{ \left| {{g_{li}}} \right|\},
\end{equation}
\begin{equation}\label{variance_1}
{\sigma ^2} = \bar N({\mathbb E}\{ {\left| {{f_{li}}} \right|^2}\} {\mathbb E}\{ {\left| {{g_{li}}} \right|^2}\}  - {({\mathbb E}\{ \left| {{f_{li}}} \right|\} )^2}{({\mathbb E}\{ \left| {{g_{li}}} \right|\} )^2})   
\end{equation}
with
\begin{equation*}
\begin{aligned}
{\mathbb E}\{ {\left| {{f_{li}}} \right|^k}\}  =& {(2{b_1})^{k/2}}\exp \left( { - \frac{{a_1^2}}{{2{b_1}}}} \right)\Gamma \left( {1 + \frac{k}{2}} \right)\\
&\times {}_1{F_1}\left( {1 + \frac{k}{2};1;\frac{{a_1^2}}{{2{b_1}}}} \right)
\end{aligned}    
\end{equation*}
\begin{equation*}
\begin{aligned}
{\mathbb E}\{ {\left| {{g_{li}}} \right|^k}\}  =& {(2{b_2})^{k/2}}\exp \left( { - \frac{{a_2^2}}{{2{b_2}}}} \right)\Gamma \left( {1 + \frac{k}{2}} \right)\\
&\times {}_1{F_1}\left( {1 + \frac{k}{2};1;\frac{{a_2^2}}{{2{b_2}}}} \right)
\end{aligned}    
\end{equation*}
\begin{equation*}
\begin{array}{*{20}{l}}
{{a_1} = \sqrt {\frac{{{\rho _1}{K_1}}}{{1 + {K_1}}}} ,}&{{b_1} = \frac{{{\rho _1}}}{{2(1 + {K_1})}}}\\
{{a_2} = \sqrt {\frac{{{\rho _2}{K_2}}}{{1 + {K_2}}}} ,}&{{b_2} = \frac{{{\rho _2}}}{{2(1 + {K_2})}}}
\end{array} 
\end{equation*}
for $k=1,2$. Moreover, $w \sim {\cal{CN}}(0,\sigma _w^2)$ with $\sigma _w^2=N\rho _2\sigma _r^2( {\sum\nolimits_{l = 1}^L {({{\bf{a}}_k(l)})^2} })/L + \sigma _0^2$.
\end{comment}

The maximum likelihood (ML) detector is adopted to jointly detect the AAP and modulated symbol, which can be formulated as
\begin{comment}
\begin{equation}
(\hat s,\hat {\cal A}) = \mathop {\arg \min }\limits_{s,{a_l} \in {\cal A}} {\left| {y - \sqrt {{P_t}} \left( {\sum\limits_{l = 1}^L {\sum\limits_{i = 1}^{\bar N} {{a_l}{h_{li}}{g_{li}}\exp (j{\theta _{li}})} } } \right)s} \right|^2}
\end{equation}
\end{comment}
\begin{equation}
(\hat m,\hat {k}) = \mathop {\arg \min }\limits_{s_m\in {\cal S},{\bf{a}}_k\in {\cal A}} {\left| {y - \sqrt {{P_{{\rm{AP}}}}} {{\bf{h}}^T}{\bf{a}}_ks_m} \right|^2}.
\end{equation}

\subsection{Power Consumption Model}
According to the widely used amplifier power consumption model in the literature \cite{long2021active}, the total power consumed by the proposed ARIS-ADRM scheme is given by
\begin{equation}\label{PC_1}
P_{\rm {ARIS-ADRM}} = {P_{\rm{AP}}} + \eta {P_{\rm {a}}} + N{P_{\rm c}} + L{P_{\rm{DC}}},
\end{equation}
where $P_{\rm {a}}$ is the output power of the active RIS, $\eta$ is the amplifier efficiency, $P_c$ and $P_{\rm{DC}}$ are the power consumption due to 
switch and control circuits, and the direct current biasing power of each RE, respectively. The output power of the active RIS can be obtained from \eqref{received_signal} as
\begin{equation}
\begin{aligned}
{P_{\rm{a}}} &= {P_{{\rm{AP}}}}{\left| {\left( {\sum\limits_{l = 1}^L {\sum\limits_{i = 1}^{\bar N} {{\bf a}_k(l){f_{li}}\exp (j{\theta _{li}})} } } \right)} \right|^2}\\
&\qquad + \sigma _r^2{\left| {\sum\limits_{l = 1}^L {\sum\limits_{i = 1}^{\bar N} {{\bf a}_k(l)\exp (j{\theta _{li}})} } } \right|^2}\\
&= {P_{{\rm{AP}}}}{\left| {{{\mathbf{p}}^H}{{\bf{a}}_k}} \right|^2} + \sigma _r^2{\left| {{{\bf{a}}_k}} \right|^2},
\end{aligned}    
\end{equation}
where ${\mathbf{p}} = {[{p _1},...,{p _l},...,{p _L}]^T}$ with ${p _l} = \sum\limits_{i = 1}^{\bar N} {{f_{li}}\exp (j{\theta _{li}})}$.

Since the value of $P_{\rm {DC}}$ and the number of $L$ is very small, the last term in \eqref{PC_1} can be omitted \cite{zhi2022active}. In this work, we set $\eta=1$ to obtain the maximum amplifier efficiency \cite{sanila2023joint}. Thus, \eqref{PC_1} can be further simplified as
\begin{equation}\label{Power_1}
 P_{\rm {ARIS-ADRM}} = {P_{\rm{AP}}} + {P_{\rm {a}}} + N{P_{\rm c}}.   
\end{equation}
By contrast, the total power consumed by the passive RIS-assisted phase-domain reflection modulation (PRIS-PDRM) \cite{guo2020reflecting} is given by
\begin{equation}\label{Power_2}
 P_{\rm {PRIS-PDRM}} = {P_{\rm{AP}}} + N{P_{\rm c}}.   
\end{equation}
From \eqref{Power_1} and \eqref{Power_2}, it is clear that active RIS requires additional power for amplification and biasing, leading to higher overall power consumption compared to passive RIS. However, this increased power consumption allows the ARIS-ADRM scheme to achieve enhanced control over both amplitude and phase. This capability results in substantial performance improvements in terms of spectral efficiency and system adaptability, as demonstrated by the simulation results in Section VI.

\textit{Remark 1}: Active RIS is distinct from relay-type RISs with RF components \cite{he2021channel,nguyen2022hybrid,basar2019transmission} and traditional relaying technologies like amplify-and-forward (AF) or decode-and-forward (DF) relays \cite{ntontin2019multi}. Relay-type RISs use active RF chains for baseband processing and pilot transmission, while active RIS focuses solely on reflecting and amplifying signals, eliminating complex processing and minimizing hardware complexity. Unlike relays that require two time slots for signal processing and retransmission, active RIS performs amplification and reflection within a single time slot, reducing latency and enhancing spectral efficiency. Full-duplex amplify-and-forward (FD-AF) relays introduce delays due to inter-symbol dependency and require advanced receiver-side processing for optimal performance \cite{ntontin2019multi}. In contrast, active RIS amplifies and reflects signals directly, avoiding delays and enabling efficient, low-latency, high-throughput operation. The adoption of active RIS in our proposed IM scheme leverages its advantages over smart repeaters and relaying technologies. Its low-latency, high-efficiency operation and simpler, cost-effective architecture make it ideal for modern wireless systems, meeting the scheme's design goals of high spectral efficiency and energy efficiency.

\subsection{Implementation Challenges of The Proposed ARIS-ADRM System}
In this subsection, we discuss several practical challenges involved in implementing the proposed system, focusing on channel estimation, hardware complexity and power efficiency.

\textit{Channel estimation:} Channel estimation is a critical yet complex problem that has been extensively explored in the literature\cite{chen2023channel,schroeder2022two,yang2023active}. For instance, a two-stage channel estimation method was introduced in \cite{chen2023channel}, while the authors of \cite{schroeder2022two} developed a least squares (LS)-based channel estimator and optimized the training reflection patterns of active RIS. Furthermore, \cite{yang2023active} proposed an estimation protocol for active RIS systems that minimizes hardware costs and pilot overhead. Leveraging channel reciprocity and a time-division duplex protocol, the cascaded active RIS channels can be efficiently estimated at the AP using the cascaded channel estimation methods outlined in \cite{chen2023channel,schroeder2022two,yang2023active}.

\textit{Hardware implementation:} The key component of an active RIS element is the additionally integrated active reflection-type amplifier, which can be realized by different existing active components, such current-inverting converters \cite{lonvcar2019ultrathin}, asymmetric current mirrors\cite{bousquet20114}, or some integrated circuits\cite{kishor2011amplifying}. However, the integration and deployment of these components pose several hardware challenges, including efficient thermal management, seamless component integration, and cost-effective manufacturing. Addressing these challenges is critical to ensuring the practicality and scalability of active RIS systems in real-world applications.

\textit{Power efficiency:} Active RIS systems outperform passive RIS by amplifying signals, offering higher SNR, extended communication ranges, and enhanced reliability \cite{zhi2022active}. However, these benefits come with increased power consumption due to the active components like amplifiers, which contrasts with the negligible power usage of passive RIS. To solve this challenges, several strategies can be employed to improve power efficiency: 1) Energy harvesting: incorporating energy harvesting techniques, such as solar panels or RF energy scavenging, to supply power to active components and reduce dependence on external power sources \cite{li2024intelligent}. 2) Dynamic power management: implementing intelligent power management systems that adaptively allocate power to RIS elements based on operational requirements, ensuring efficient use of available energy resources \cite{zhi2022active}. 3) Low-power component design: developing energy-efficient active components, such as low-power amplifiers and optimized circuits, to minimize energy consumption while maintaining performance \cite{wu2019towards}.

\section{Performance Analysis}
In this section, we conduct a comprehensive performance analysis of the proposed scheme, focusing on achievable rate and ABEP. Detailed results will be presented in the subsequent subsections.

\subsection{Achievable Rate Performance Analysis}
In this subsection, we analyze the achievable rate of the proposed ARIS-ADRM scheme from an information-theoretic perspective. Our primary objective is to derive the mutual information (MI) between the transmitted and received signals. By determining the MI, we can precisely quantify the maximum achievable data rate under a variety of transmission conditions. This approach allows us to rigorously assess the performance limits of the ARIS-ADRM scheme, providing a comprehensive understanding of its efficiency and potential in different scenarios. 

To simplify the notation, let us denote ${\bf x}_{\xi}={\bf a}_ks_m$, where $\xi=1,2,...,2^R$. The probability density function (PDF) of the received signal $y$ given the input $\bf x$ and channel $\bf h$ is expressed as
\begin{equation}
p(y\left| {\bf{x}} \right.,{\bf{h}}) = \frac{1}{{\pi \sigma _w^2}}\exp \left( { - \frac{{{{\left| {y - \sqrt {{P_{{{\rm {AP}}}}}} {{\bf{h}}^T}{\bf{x}}} \right|}^2}}}{{\sigma _w^2}}} \right).
\end{equation}
Therefore, the PDF of the signal $y$ given the channel $\bf h$ can be written as
\begin{equation}
p(y\left| {\bf{h}} \right.) = {{\mathbb E}_{\bf{x}}}\left\{ {p(y\left| {\bf{x}} \right.,{\bf{h}})} \right\} = \frac{1}{{{2^R}}}\sum\limits_{\xi  = 1}^{{2^R}} {p(y\left| {{{\bf{x}}_\xi }} \right.,{\bf{h}})}.
\end{equation}

The MI between the input $\bf x$ and the channel output $y$ in the proposed ARIS-ADRM scheme can be calculated by
\begin{equation}\label{MI_1}
\begin{aligned}
I({\bf{x}};y\left| {\bf{h}} \right.)& = {\cal H}({\bf{x}}) - {\cal H}({\bf{x}}\left| {y,{\bf{h}}} \right.) \\
&= R - \sum\limits_{\xi  = 1}^{{2^R}} {\int_y {p({\bf{x}} = {{\bf{x}}_\xi },y\left| {\bf{h}} \right.)} }  \times \\
&\qquad {\log _2}\frac{{p\left( {y\left| {\bf{h}} \right.} \right)}}{{p\left( {{\bf{x}} = {{\bf{x}}_\xi }} \right)p\left( {y\left| {{{\bf{x}}_\xi },{\bf{h}}} \right.} \right)}}{\rm{d}}y\\
& = R - \sum\limits_{\xi  = 1}^{{2^R}} {\int_y {p({\bf{x}} = {{\bf{x}}_\xi })p(y\left| {{{\bf{x}}_\xi },} \right.{\bf{h}})} }  \times \\
&\qquad {\log _2}\frac{{p\left( {y\left| {\bf{h}} \right.} \right)}}{{p\left( {{\bf{x}} = {{\bf{x}}_\xi }} \right)p\left( {y\left| {{{\bf{x}}_\xi },{\bf{h}}} \right.} \right)}}{\rm{d}}y.
\end{aligned}  
\end{equation}
When the signal vector ${\bf x}_{\xi}$ is transmitted, the received signal is given by $y={\bf h}^T{\bf x}_{\xi}+{w}$. In this scenario, the only unknown in $y$ is the noise $w$. Consequently, the integral over $ y$ can be transformed into an integral over $w$. 

By substituting ${y}-{\bf h}^T{\bf x}_{\xi}={w}$ and $p\left( {{\bf{x}} = {{\bf{x}}_\xi}} \right) = \frac{1}{{{2^R}}}$ into  the $\xi$th integral of the summation in \eqref{MI_1}, we obtain
\begin{equation}\label{MI_2}
\begin{aligned}
&\int_y {p\left( {{\bf{x}} = {{\bf{x}}_\xi }} \right)p\left( {y\left| {{{\bf{x}}_\xi },{\bf{h}}} \right.} \right){{\log }_2}\frac{{p\left( {y\left| {\bf{h}} \right.} \right)}}{{p\left( {{{\bf{x}}_\xi }} \right)p\left( {y\left| {{{\bf{x}}_\xi },{\bf{h}}} \right.} \right)}}{\rm{d}}y} \\
& = \frac{R}{{{2^R}}}\int_w {p\left( {{{\bf{h}}^T}{{\bf{x}}_\xi } + w\left| {{{\bf{h}}^T}{{\bf{x}}_\xi }} \right.} \right){{\log }_2}\frac{{p\left( {{{\bf{h}}^T}{{\bf{x}}_\xi } + w} \right)}}{{p\left( {{{\bf{h}}^T}{{\bf{x}}_\xi } + w\left| {{{\bf{h}}^T}{{\bf{x}}_\xi }} \right.} \right)}}{\rm{d}}} w\\
& = \frac{R}{{{2^R}}}\int_w {p\left( w \right){{\log }_2}\frac{{p\left( {{{\bf{h}}^T}{{\bf{x}}_\xi } + w} \right)}}{{p\left( w \right)}}{\rm{d}}w} \\
& = \frac{R}{{{2^R}}}\int_w {p\left( w \right){{\log }_2}\frac{{\sum\limits_{\hat \xi  = 1}^{{2^R}} {p\left( {{\bf{x}} = {{\bf{x}}_{\hat \xi }}} \right)p\left( {{{\bf{h}}^T}{{\bf{x}}_\xi } + w\left| {{{\bf{h}}^T}{{\bf{x}}_{\hat \xi }}} \right.} \right)} }}{{p\left( w \right)}}{\rm{d}}w} \\
& = \frac{1}{{{2^R}}}\int_w {p\left( w \right){{\log }_2}\frac{{\sum\limits_{\hat \xi  = 1}^{{2^R}} {p\left( {{{\bf{h}}^T}{{\bf{x}}_\xi } + w\left| {{{\bf{h}}^T}{{\bf{x}}_{\hat \xi }}} \right.} \right)} }}{{p\left( w \right)}}{\rm{d}}w} 
\end{aligned}
\end{equation}
where the integrals and expectations are taken over the noise $w$. The expressions for the probability densities are given by
\begin{equation}\label{MI_3}
\left\{ {\begin{array}{*{20}{l}}
{p\left( {{{\bf{h}}^T}{{\bf{x}}_\xi } + w\left| {{{\bf{h}}^T}{{\bf{x}}_{\hat \xi }}} \right.} \right) = \frac{1}{{\pi \sigma _w^2}}\exp \left( {\frac{{ - {{\left| {({{\bf{h}}^T}{{\bf{x}}_\xi } - {{\bf{h}}^T}{{\bf{x}}_{\hat \xi }}) + w} \right|}^2}}}{{\sigma _w^2}}} \right),}\\
{p\left( w \right) = \frac{1}{{\pi \sigma _w^2}}\exp \left( { - \frac{{{{\left| w \right|}^2}}}{{\sigma _w^2}}} \right).}
\end{array}} \right..
\end{equation}

Substituting \eqref{MI_2} and \eqref{MI_3} into \eqref{MI_1}, the MI of our proposed ARIS-ADRM system can be further formulated as
\begin{equation}\label{MI}
\begin{aligned}
I\left( {{\bf{x}};y\left| {\bf{h}} \right.} \right) &= R - \frac{1}{{{2^R}}}\sum\limits_{\xi  = 1}^{{2^R}} {\int_w {p\left( w \right){{\log }_2}\sum\limits_{\hat \xi  = 1}^{{2^R}} {\exp ( - {d_{\xi ,\hat \xi }}(w))} {\rm{d}}w} } \\
& = R - \frac{1}{{{2^R}}}\sum\limits_{\xi  = 1}^{{2^R}} {{{\mathbb E}_w}\left[ {{{\log }_2}\sum\limits_{\hat \xi  = 1}^{{2^R}} {\exp ( - {d_{\xi ,\hat \xi }}(w))} } \right]},
\end{aligned}
\end{equation}
where ${d_{\xi ,\hat \xi }}(w) = \frac{{{{\left| {({{\bf{h}}^T}{{\bf{x}}_\xi } - {{\bf{h}}^T}{{\bf{x}}_{\hat \xi }}) + w} \right|}^2} - {{\left| w \right|}^2}}}{{\sigma _w^2}}$.

While the MI expression for the proposed ARIS-ADRM scheme given in \eqref{MI} does not yield a closed-form solution, analyzing the behavior of MI in various limits can provide valuable insights into the performance boundaries of the proposed scheme. In particular, in the limit of very low SNR region (i.e., SNR $\gamma  = \frac{P_T}{{{\sigma_w ^2}}} \to 0$), we have
\begin{equation}
\left\{ {\begin{array}{*{20}{l}}
{{{\left| {({{\bf{h}}^T}{{\bf{x}}_\xi } - {{\bf{h}}^T}{{\bf{x}}_{\hat \xi }}) + w} \right|}^2} \to {{\left| w \right|}^2},\xi  \ne \hat \xi }\\
{{{\left| {({{\bf{h}}^T}{{\bf{x}}_\xi } - {{\bf{h}}^T}{{\bf{x}}_{\hat \xi }}) + w} \right|}^2} = {{\left| w \right|}^2},\xi  = \hat \xi }
\end{array}} \right.,
\end{equation}
where $P_T$ represents the total transmit power in the scheme, therefore, ${d_{\xi ,\hat \xi }}(w)=0$ in the low SNR region. As a result, the limit of \eqref{MI} can be written as (assuming a fixed noise power ${{\sigma_w ^2}}=1$)
\begin{equation}\label{limit_1}
\begin{aligned}
&\mathop {\lim }\limits_{\gamma  \to 0} I\left( {{\bf{x}};y\left| {\bf{h}} \right.} \right) = \mathop {\lim }\limits_{\scriptstyle{P_T} \to 0\hfill\atop
\scriptstyle\sigma _w^2 = 1\hfill} I\left( {{\bf{x}};y\left| {\bf{h}} \right.} \right)\\
& = R - \mathop {\lim }\limits_{\scriptstyle{P_T} \to 0\hfill\atop
\scriptstyle\sigma _w^2 = 1\hfill} \frac{1}{{{2^R}}}\sum\limits_{\xi  = 1}^{{2^R}} {{{\mathbb E}_w}\left[ {{{\log }_2}\sum\limits_{\hat \xi  = 1}^{{2^R}} {\exp ( - {d_{\xi ,\hat \xi }}(w))} } \right]} \\
& = 0.
\end{aligned}
\end{equation}

On the other hand, in the limit of very high SNR region (i.e., $\gamma  = \frac{P_T}{{{\sigma_w ^2}}} \to +{\infty}$), the limit of \eqref{MI} is
\begin{equation}\label{limit_2}
\begin{aligned}
&\mathop {\lim }\limits_{\gamma  \to  + \infty } I\left( {{\bf{x}};y\left| {\bf{h}} \right.} \right) = \mathop {\lim }\limits_{\scriptstyle{P_T} \to  + \infty \hfill\atop
\scriptstyle\sigma _w^2 = 1\hfill} I\left( {{\bf{x}};y\left| {\bf{h}} \right.} \right)\\
& = \mathop {\lim }\limits_{\scriptstyle\xi  \ne \hat \xi ,{\left| {({{\bf{h}}^T}{{\bf{x}}_\xi } - {{\bf{h}}^T}{{\bf{x}}_{\hat \xi }}) + w} \right|^2} \to \infty \hfill\atop
{\scriptstyle\xi  = \hat \xi ,{\left| {({{\bf{h}}^T}{{\bf{x}}_\xi } - {{\bf{h}}^T}{{\bf{x}}_{\hat \xi }}) + w} \right|^2} = {\left| w \right|^2}\hfill\atop
\scriptstyle\sigma _w^2 = 1\hfill}} I\left( {{\bf{x}};y\left| {\bf{h}} \right.} \right)\\
& = R.
\end{aligned}
\end{equation}
The limits in \eqref{limit_1} and \eqref{limit_2} demonstrate fundamental achievable rate boundaries of the proposed ARIS-ADRM scheme under different SNR conditions, highlighting its 
sensitivity to transmit power and noise levels in achieving efficient information transmission.

\subsection{BER Performance Analysis}
The conditional pair-wise error probability (CPEP) of the proposed scheme can be written as
\begin{equation}
\begin{aligned}
&{\mathbb P}\left\{ {(m,k) \to ({{\hat m}},\hat k)\left| {\bf{h}} \right.} \right\} \\
&= {\mathbb P}\left\{ {\left| {y - \sqrt {{P_{{\rm{AP}}}}} {{\bf{h}}^T}{\bf{a}}_ks_m} \right|^2} \right. \left. { > \left| {y - \sqrt {{P_{{\rm{AP}}}}} {{\bf{h}}^T}{{ {\bf a}_{\hat k}}}{s}_{\hat m}} \right|^2} \right\}\\
& = {\mathbb P}\{\zeta<0\},
\end{aligned}
\end{equation}
where
\begin{equation}
\begin{aligned}
\zeta = &{\left| {\sqrt {{P_{{\rm{AP}}}}} {{\bf{h}}^T}({\bf{a}}_ks_m - {\bf{a}}_{\hat k} s_{\hat m})} \right|^2} \\
&\qquad + 2\Re \{ \sqrt {{P_{{\rm{AP}}}}} {w^*}{{\bf{h}}^T}({\bf{a}}_ks_m - {\bf{a}}_{\hat k}s_{\hat m})\}.
\end{aligned}
\end{equation}
Let $\Delta  =  {{\bf{h}}^T}({\bf{a}}_ks_m - {\bf{ a}}_{\hat k} s_{\hat m})$ for ease of notation, thus $\zeta$ can be rewritten as $\zeta  = {{P_{{\rm{AP}}}}}{\left| \Delta  \right|^2} + 2\Re \{ \sqrt {{P_{{\rm{AP}}}}}{w^*}\Delta \}$, which is a Gaussian random variable with mean $\mu_{\zeta}$ and variance $\sigma_{\zeta}^2$, i.e., $\zeta \sim {\cal N}(\mu_{\zeta},\sigma_{\zeta}^2)$, where
\begin{equation}
{\mu _\zeta } = {\mathbb E}[{{P_{{\rm{AP}}}}}{\left| \Delta  \right|^2} + 2\Re \{ \sqrt {{P_{{\rm{AP}}}}}{w^*}\Delta \} ] = {{P_{{\rm{AP}}}}}{\left| \Delta  \right|^2},  
\end{equation}
and
\begin{equation}
\begin{aligned}
\sigma _\zeta ^2 &= {\mathbb E}[{(\zeta  - {\mu _\zeta })^2}]\\
& = {\mathbb E}[{(2\Re \{\sqrt {{P_{{\rm{AP}}}}} {w^*}\Delta \} )^2}]\\
& = {\mathbb E}[4 {{P_{{\rm{AP}}}}}\Re \{ {({w^*})^2}\} ]{\mathbb E}[\Re \{ {\left| \Delta  \right|^2}\} ]\\
& = 2{{P_{{\rm{AP}}}}}\sigma _w^2{\left| \Delta  \right|^2}.
\end{aligned}    
\end{equation}
Therefore, the CPEP can be further obtained as
\begin{equation}\label{CPEP_1}
{\mathbb P}\left\{ {(m,k) \to ({{\hat m}},\hat k)\left| {\bf{h}} \right.} \right\} = Q\left( {\sqrt {\frac{{{{P_{{\rm{AP}}}}}{{\left| \Delta  \right|}^2}}}{{2\sigma _w^2}}} } \right),
\end{equation}
where $Q(x)$ is the Gaussian $Q$-function. By approximating the $Q$-function as $Q(x)\approx \frac{1}{{12}}\exp ( - {{{x^2}} \mathord{\left/
 {\vphantom {{{x^2}} 2}} \right.
 \kern-\nulldelimiterspace} 2}) + \frac{1}{4}\exp ( - {{2{x^2}} \mathord{\left/
 {\vphantom {{2{x^2}} 3}} \right.
 \kern-\nulldelimiterspace} 3})$, the CPEP can be approximated as
 \begin{equation}\label{CPEP_2}
 \begin{aligned}
& {\mathbb P}\left\{ {(m,k) \to ({{\hat m}},\hat k)\left| {\bf{h}} \right.} \right\} \\
&\quad \approx \frac{1}{{12}}\exp \left( { - \frac{{ {P_{\rm {AP}}}{{\left| \Delta  \right|}^2}}}{{4\sigma _w^2}}} \right) + \frac{1}{4}\exp \left( { - \frac{{{P_{\rm{AP}}}{{\left| \Delta  \right|}^2}}}{{3\sigma _w^2}}} \right).  
\end{aligned}
 \end{equation}
Thus, the unconditional PEP (UPEP) can be obtained by averaging \eqref{CPEP_2} over $\bf h$ as
\begin{equation}
\begin{aligned}
&{\mathbb P}\left\{ {(m,k) \to ({{\hat m}},\hat k)} \right\}\\
& \approx {{\mathbb E}_\Delta }\left[ {\frac{1}{{12}}\exp \left( { - \frac{{{{P_{{\rm{AP}}}}}{{\left| \Delta  \right|}^2}}}{{4\sigma _w^2}}} \right)} \right] + {{\mathbb E}_\Delta }\left[ {\frac{1}{4}\exp \left( { - \frac{{{P_{\rm{AP}}}{{\left| \Delta  \right|}^2}}}{{3\sigma _w^2}}} \right)} \right]\\
& = \frac{1}{{12}}{M_\Delta }\left( { - \frac{{{{P_{\rm{AP}}}}}}{{4\sigma _w^2}}} \right) + \frac{1}{4}{M_\Delta }\left( { - \frac{{{P_{\rm{AP}}}}}{{3\sigma _w^2}}} \right),  
\end{aligned}    
\end{equation}
where $M_\Delta(t)={{\mathbb E}_\Delta}[\exp(t{{{\left| \Delta  \right|}^2}})]$ is the moment-generating function (MGF) of ${{{\left| \Delta  \right|}^2}}$. According to the CLT, when $N \gg 1$,  $h_l=\sum\nolimits_{i = 1}^{\bar N} {\left| {{f_{li}}} \right|\left| {{g_{li}}} \right|}\sim {\cal N}(\mu,\sigma^2)$, where the derivations of $\mu$ and $\sigma^2$ are detailed in Appendix B. As a result, ${{{\left| \Delta  \right|}^2}}$ follows the non-central Chi-square distribution with two degrees of freedom. The MGF of ${{{\left| \Delta  \right|}^2}}$ is given by
\begin{equation}
{M_\Delta }(t) = \frac{1}{\sqrt{1 - 2\beta {\sigma ^2}t}}\exp \left( {\frac{{{\mu ^2}\beta t}}{{1 - 2\beta {\sigma ^2}t}}} \right),
\end{equation}
where $\beta  = {\left| {{{\bf{a}}_k}{s_m} - {{\bf{a}}_{\hat k}}{s_{\hat m}}} \right|^2}$.

\begin{comment}
According to the CLT, when $N \gg 1$,  $h_l=\sum\nolimits_{i = 1}^{\bar N} {\left| {{f_{li}}} \right|\left| {{g_{li}}} \right|}\sim {\cal N}(\mu,\sigma^2)$, where
\begin{equation}\label{mean_1}
\mu  = \bar N{\mathbb E}\{ \left| {{f_{li}}} \right|\} {\mathbb E}\{ \left| {{g_{li}}} \right|\},
\end{equation}
\begin{equation}\label{variance_1}
{\sigma ^2} = \bar N({\mathbb E}\{ {\left| {{f_{li}}} \right|^2}\} {\mathbb E}\{ {\left| {{g_{li}}} \right|^2}\}  - {({\mathbb E}\{ \left| {{f_{li}}} \right|\} )^2}{({\mathbb E}\{ \left| {{g_{li}}} \right|\} )^2})   
\end{equation}
with
\begin{equation*}
\begin{aligned}
{\mathbb E}\{ {\left| {{f_{li}}} \right|^k}\}  =& {(2{b_1})^{k/2}}\exp \left( { - \frac{{a_1^2}}{{2{b_1}}}} \right)\Gamma \left( {1 + \frac{k}{2}} \right)\\
&\times {}_1{F_1}\left( {1 + \frac{k}{2};1;\frac{{a_1^2}}{{2{b_1}}}} \right)
\end{aligned}    
\end{equation*}
\begin{equation*}
\begin{aligned}
{\mathbb E}\{ {\left| {{g_{li}}} \right|^k}\}  =& {(2{b_2})^{k/2}}\exp \left( { - \frac{{a_2^2}}{{2{b_2}}}} \right)\Gamma \left( {1 + \frac{k}{2}} \right)\\
&\times {}_1{F_1}\left( {1 + \frac{k}{2};1;\frac{{a_2^2}}{{2{b_2}}}} \right)
\end{aligned}    
\end{equation*}
\begin{equation*}
\begin{array}{*{20}{l}}
{{a_1} = \sqrt {\frac{{{\rho _1}{K_1}}}{{1 + {K_1}}}} ,}&{{b_1} = \frac{{{\rho _1}}}{{2(1 + {K_1})}}}\\
{{a_2} = \sqrt {\frac{{{\rho _2}{K_2}}}{{1 + {K_2}}}} ,}&{{b_2} = \frac{{{\rho _2}}}{{2(1 + {K_2})}}}
\end{array} 
\end{equation*}
for $k=1,2$.
\end{comment}

Finally, the ABEP of the proposed ARIS-ADRM can be analyzed to be upper bounded by
\begin{equation}\label{BER_final}
\begin{aligned}
{\rm {ABEP}}  \le {{\bar {\mathbb P}}_e} = &\frac{1}{{R{2^R}}}\sum\limits_m {\sum\limits_k {\sum\limits_{\hat m} {\sum\limits_{\hat k} {d\{ (m,k) \to (\hat m,\hat k)\} } } } } \\
&\times {\mathbb P}\{(m,k) \to (\hat m,\hat k)\},
\end{aligned}
\end{equation}
where $R$ is the transmission rate given in \eqref{Data_rate}, ${d\{ (m,k) \to (\hat m,\hat k)\} }$ is the Hamming distance. 

\textit{Remark 2:} In our study, MI analysis provides insights into the spectral efficiency and theoretical performance limits of the proposed scheme, while CPEP analysis reflects the practical error performance under specific channel conditions. The interconnection arises from the fact that a lower CPEP generally corresponds to higher MI, as reduced error probabilities signify improved signal detection reliability, enabling the system to operate closer to its theoretical capacity.\\

\section{Amplitude Allocation Pattern Codebook Design}
In this work, one key concern is how to design the AAP codebook $\cal A$. In this section, we will explore the AAP codebook design with the objective of minimizing the ABEP upper bound ${{{\overline {\mathbb P} }_e}}$, assuming the availability of perfect CSI at the transmitter. The corresponding optimization problem can be formulated as
\begin{align}\label{P1}
&\mathop {\min }\limits_{{\bf a}_k} \quad {{{\overline {\mathbb P} }_e}} \\
&{\rm{s.t.}}\quad \, {{P_{{\rm{AP}}}}{{\left| {{{\bf{p}}^H}{{\bf{a}}_k}} \right|}^2} + \sigma _r^2{{\left| {{{\bf{a}}_k}} \right|}^2} \le {P_{\rm{a}}}},\forall k = 1,..,A, \tag{32a}\\
&\qquad \,\,\, 1<{{{\bf{a}}_k}(l) \le {\alpha _{{\rm{max}}}},\forall l = 1,..,L},\tag{32b}
\end{align}
where (32a) is the total power constraint at the active RIS and (32b) is the power constraint for each RE. Despite the convex nature of these constraints, the optimization problem remains challenging due to the non-convexity  of the objective function.
According to \eqref{BER_final}, minimizing ${{{\overline {\mathbb P} }_e}}$ is equivalent to 
minimizing the UPEP ${\mathbb P}\{(m,k) \to (\hat m,\hat k)\}$, which is further equivalent to maximizing the minimum squared Euclidean distance ${\left| \Delta  \right|^2}$ \cite{yang2019enhanced}. As a result, the optimization problem in \eqref{P1} can be further converted to maximization of the minimum squared Euclidean distance problem as
\begin{align}\label{P11}
&\mathop {\max }\limits_{{\bf a}_k} \quad {\min {{\left| \Delta  \right|}^2}}\\
&{\rm{s.t.}} \qquad  {(32\rm{a})},{(32\rm{b})}\nonumber.
\end{align}
To solve this problem, we introduce the following new matrix and vectors, 
\begin{align}\label{SCA1}
&{\bf{D}}_{\bf a} = \left[ {\begin{array}{*{20}{l}}
{{{\bf{A}}_1}}&{\bf{0}}& \cdots &{\bf{0}}\\
{\bf{0}}&{{{\bf{A}}_2}}& \cdots &{\bf{0}}\\
 \vdots & \vdots & \ddots & \vdots \\
{\bf{0}}&{\bf{0}}& \cdots &{{{\bf{A}}_A}}
\end{array}} \right] \in {{\mathbb R}^{AL \times AL}}, \\
&\tilde{\bf{h}} = [\underbrace {{{\bf{h}}^T},{{\bf{h}}^T},...,{{\bf{h}}^T}}_A] \in {{\mathbb C}^{1 \times AL}},\\
& {\bf s}_m = [\underbrace{{{s_m}},{s_m},...,{s_m}}_L]^T \in {{\mathbb C}^{L \times 1}},\\
& {{\bf{\Psi}}_{q}} = {{\bf{e}}_k} \otimes {{\bf{s}}_m} \in {{\mathbb C}^{AL \times 1}},
\end{align}
where ${\bf{A}}_k ={\rm{diag(}}{{\bf{a}}_k}{\rm{)}}$ and ${\bf e}_k \in {\mathbb R}^{A \times 1}$ is the $k$th column of the identity matrix ${\bf I}_A$,$\forall k=1,2,...,A$ and $q=1,...,2^R$. As a result, the square of Euclidean distances can be rewritten as
\begin{equation}\label{SCA2}
\begin{aligned}
{\left| \Delta  \right|^2} &= {\left| {{{\bf{h}}^T}{{\bf{a}}_k}{s_m} - {{\bf{h}}^T}{{\bf{a}}_{\hat k}}{s_{\hat m}}} \right|^2}\\
& = {\left| {\bf{\tilde h}}{{\bf{D}}_{\bf a}({{\bf{\Psi}}_q} - {{\bf{\Psi}}_{\hat q}})} \right|^2}\\
& = {({{\bf{\Psi}}_q} - {{\bf{\Psi}}_{\hat q}})^H}{\bf{D}}_{\bf a}^H{{{\bf{\tilde h}}}^H}{\bf{\tilde h}}{\bf{D}}_{\bf a}({{\bf{\Psi}}_q} - {{\bf{\Psi}}_{\hat q}})\\
& = {\rm{Tr(}}{\bf{D}}_{\bf a}^H{\bf{D}}_{\bf a}{{{\bf{\tilde h}}}^H}{\bf{\tilde h}}({{\bf{\Psi}}_q} - {{\bf{\Psi}}_{\hat q}}){({{\bf{\Psi}}_q} - {{\bf{\Psi}}_{\hat q}})^H}{\rm{)}}\\
& = {{\bf{a}}^T}{{\bf{R}}_{q,\hat q}}{\bf{a}},
\end{aligned}
\end{equation}
where ${\bf a}=[{\bf a}_1^T,...,{\bf a}_k^T,...,{\bf a}_A^T]^T={\rm{diag}}({\bf D}_{\bf a})\in {\mathbb R}^{AL\times 1}$ and ${{\bf{R}}_{q,\hat q}}={{{\bf{\tilde h}}}^H}{\bf{\tilde h}} \odot [({{\bf{\Psi}}_q} - {{\bf{\Psi}}_{\hat q}}){({{\bf{\Psi}}_q} - {{\bf{\Psi}}_{\hat q}})^H}]\in{\mathbb C}^{AL \times AL}$. 
Note that the instantaneous CSI is encapsulated within the matrix ${\bf R}_{q,\hat q}$.
Furthermore, the optimization problem in \eqref{P11} can be recast into the following problem 
\begin{align}\label{P2}
&\mathop {\max }\limits_{{\bf a}_k} \quad {\rm {min}}\,{{\bf{a}}^T}{{\bf{R}}_{q,\hat q}}{\bf{a}}, \forall q, \hat q,q \ne \hat q,\\
&{\rm{s.t.}} \qquad {{\bf{a}}_k^T}{\bf{Fa}}_k \le {P_{\rm{a}}},\forall k=1,..,A,\tag{39a}\\
&\qquad \,\, \quad 1 < {{{\bf{a}}}(i) \le {\alpha _{{\rm{max}}}},\forall i = 1,..,AL}.\tag{39b}
\end{align}
where ${\bf{F}} = {P_{{\rm{AP}}}}{\rm{diag(}}{\left|{p_1}\right|}^2,...,{\left|{p_l}\right|}^2,...,{\left|{p_L}\right|}^2{\rm{)}} + \sigma _r^2{{\bf{I}}_L}$. It is worth noting that the optimization problem \eqref{P2} is a non-convex QCQP, since the objective function ${{\bf{a}}^T}{{\bf{R}}_{q,\hat q}}{\bf{a}}$ is a non-convex quadratic function.

To tackle this problem effectively, we first introduce an auxiliary scalar variable $\tau$ to reformulate \eqref{P2} into a more tractable form as
\begin{align}\label{P3}
&\mathop {\max }\limits_{{\bf a}_k} \quad \tau \\
&{\rm{s.t.}} \qquad {{\bf{a}}^T}{{\bf{R}}_{q,\hat q}}{\bf{a}} \ge \tau, \forall q, \hat q,q \ne \hat q,\tag{40a}\\
&\qquad \,\, \quad  {{\bf{a}}_k^T}{\bf{Fa}}_k \le {P_{\rm{a}}},\forall k=1,..,A,\tag{40b}\\
&\qquad \,\, \quad 1 < {{{\bf{a}}}(i) \le {\alpha _{{\rm{max}}}},\forall i = 1,..,AL}.\tag{40c}
\end{align}
This reformulation introduces an optimization problem where the objective is to maximize $\tau$, subject to non-convex quadratic constraint (40a), linear power constrain (40b), and box constraint (40c) on the AAP codebook vector ${\bf a}_k$.
 
To proceed, we apply SCA to approximate the non-convex quadratic constraint (40a) by its first order Taylor expansion. This allows us to transform problem \eqref{P3} into an approximate SCA problem as
\begin{align}\label{P4}
&\mathop {\max }\limits_{{\bf a}_k} \quad \tau \\
&{\rm{s.t.}} \qquad \Re \left\{ {2{\bf{a}}_{(t)}^T{{\bf{R}}_{q,\hat q}}{\bf{a}} - {\bf{a}}_{(t)}^T{{\bf{R}}_{q,\hat q}}{{\bf{a}}_{(t)}}} \right\} \ge \tau, \forall q, \hat q,q \ne \hat q,\tag{41a}\\
&\qquad \,\, \quad {{\bf{a}}_k^T}{\bf{Fa}}_k \le {P_{\rm{a}}},\forall k=1,..,A,\tag{41b}\\
&\qquad \,\, \quad 1 < {{{\bf{a}}}(i) \le {\alpha _{{\rm{max}}}},\forall i = 1,..,AL},\tag{41c}
\end{align}
where ${\bf{a}}_{(t)}$ is the AAP codebook solution from the $(t-1)$th iteration. The approximation $\Re \{ {2{\bf{a}}_{(t)}^T{{\bf{R}}_{q,\hat q}}{\bf{a}} - {\bf{a}}_{(t)}^T{{\bf{R}}_{q,\hat q}}{{\bf{a}}_{(t)}}} \}$  linearizes the quadratic constraint (40a) and renders the problem convex. This enables the use of efficient optimization techniques such as the interior-point method or CVX tool to find an optimal solution. 

\begin{algorithm}[t]
\caption{Proposed SCA-based AAP Codebook Design Algorithm for the ARIS-ADRM Scheme}
\begin{algorithmic}[1]
\REQUIRE ${\bf f}, {\bf g}, M,{P_a}, {P_{\rm{AP}}}, {\alpha _{{\rm{max}}}}$\\
\ENSURE  AAP Codebook $\cal A$
\STATE Initialization: the initial solution ${\bf a}_{(0)} = {\alpha _{{\rm{max}}}}{\bf I}_{AL}$.
\STATE {\textbf{Repeat}}
\STATE \quad Update the solution ${\bf a}_{(t)}$ by solving the SCA problem \eqref{P4} with ${\bf a}_{(t-1)}$.
\STATE \textbf{Until} The value of $\tau$ converges.
\STATE Construct AAP Codebook: ${\cal A} = {\rm {reshape}}({\bf a},[L,A])$.
\end{algorithmic}
\end{algorithm}

The detailed procedure for designing the AAP codebook using SCA is outlined in \textbf{Algorithm 1}, which iteratively improves the approximation of the non-convex constraint (40a) until convergence. This approach ensures that the problem \eqref{P4} is efficiently solved while respecting the constraints on power and amplitude.

\textit{Converge Analysis:} Let ${\bf a}_{(t)}$ denote the optimal solution in the $(t-1)$th iteration of the SCA algorithm. Utilizing the linear approximation given in (41a), ${\bf a}_{(t)}$ remains feasible for the optimization problem at the $t$th iteration. As ${\bf a}_{(t+1)}$ represents the optimal solution at the $t$th iteration, it follows that $\tau_{(t+1)} \ge \tau_{(t)}$. This observation indicates that the proposed SCA algorithm generates a nondecreasing sequence. Moreover, due to the bounded nature of $\tau$ arising from the power constraint, the convergence of the proposed AAP codebook design algorithm is guaranteed. 

\begin{figure}[t]
\centering
\includegraphics[width=0.45\textwidth]{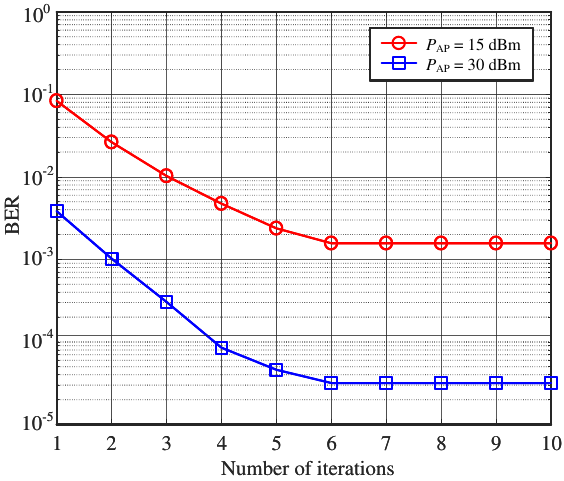}
\caption{Convergence of \textbf{Algorithm 1}.}\label{BER_Convergence}
\end{figure}

To illustrate this more intuitively, we evaluate the convergence behaviour of Algorithm 1 as shown in Fig. \ref{BER_Convergence}. The simulation parameters are set as $N=128$, $L=A=2$ and 4QAM modulation. It can be observed from the figure that the BER performance initially decreases and then converges to a fixed value, confirming the effectiveness of the proposed SCA-based AAP codebook design algorithm.

\textit{Complexity Analysis:} The complexity of the proposed SCA-based AAP codebook design algorithm can be assessed from two main perspectives as
\begin{itemize}
    \item 
    Computational Complexity: This includes the computation involved in solving the convex optimization problem at each iteration. Assuming the use of an interior point method for solving convex problems \cite{boyd2004convex}, the computational complexity for constructing problem \eqref{P4} is ${\cal O}({M^2}{A^2}L)$.
    \item
    Iteration Complexity: This refers to the number of iterations required for the algorithm to converge. The algorithm typically requires several iterations to achieve convergence, which is influenced by factors such as the initial conditions and the rate of convergence of the SCA method. A theoretical upper bound for the iteration complexity is given by ${\cal O}({M^4}{A^4})$ \cite{lee2015generalized}.
\end{itemize}

Therefore, the overall complexity of the proposed AAP codebook design algorithm, considering both the construction of the convex optimization problem and its iterative solution, is represented  by
\begin{equation}\label{Complexity}
{\cal C} = {\cal O}({M^2}{A^2}L) + {\cal O}({M^4}{A^4}).
\end{equation}
As can be seen from \eqref{Complexity}, the proposed SCA-based AAP codebook design algorithm operates efficiently within polynomial complexity bounds, demonstrating its feasibility for handling large-scale scenarios under the inherent challenges of convex optimization tasks.
\begin{comment}
\textit{Complexity Analysis:} The complexity of the proposed SCA-based AAP codebook design algorithm can be assessed from two main perspectives: 1) the computational complexity involved in solving the convex optimization problem at each iteration, and 2) the number of iterations needed to achieve the convergence. The exact computational complexity of the convex solving process depends on the specific convex solver used. For the purpose of analysis, we assume the use of the interior point method \cite{boyd2004convex}. The overall algorithm complexity is given by
\begin{equation}
{\cal C} = {\cal O}({M^2}{A^2}L) + {\cal O}({M^4}{A^4}), 
\end{equation}
where the first term ${\cal O}({M^2}{A^2}L)$ is the computational complexity for constructing the convex optimization problem \eqref{P4}, and the second term ${\cal O}({M^4}{A^4})$ is the complexity order required during the convex solving process \cite{yang2019enhanced}. 
\end{comment}

\section{Extension to MIMO Systems}
In this section, we generalize the proposed ARIS-ADRM framework to support MIMO systems and propose the ARIS-ADRM-MIMO scheme, where both the AP and user are equipped with multiple antennas. Specifically, the AP and the user have $N_t$ and $N_r$ antennas, respectively. The wireless channel between the AP and the user, the AP and the RIS, between the RIS and the user are denoted by ${\bf H} \in {\mathbb C}^{N_r \times N_t}$, ${\bf F} \in {\mathbb{C}}^{N \times N_t}$ and ${\bf G} \in {\mathbb{C}}^{N_r \times N}$, respectively. Both channel links are modeled as Rician fading, consistent with the SISO case, but with adjustments in subscripts to account for the multi-antenna configuration. The $(\tau, t)$th entry of $\bf F$, denoted by $f_{\tau t}$, follows the same formulation as in \eqref{H_1}, with the subscript $li$ replaced by $\tau t$, where $\tau = 1,...,N$ and $t = 1,..., N_t$. Similarly, the $(r, \tau)$th entry of $\bf G$, denoted by $g_{r\tau}$, is the same as \eqref{G_1}, except that the subscript $li$ is changed into $r\tau$ for $r = 1,..., N_r$. The $(r,t)$th entry of $\bf H$ is given by
\begin{equation}
h_{rt} = \sqrt {{\rho _0}} \left( {\sqrt {\frac{{{K_0}}}{{1 + {K_0}}}} {{\bar h}_{rt}} + \sqrt {\frac{1}{{1 + {K_0}}}} {{\tilde h}_{rt}}} \right),
\end{equation}
 where the components ${{\bar h}_{rt}} = \exp ( - {{j2\pi {d_0}} \mathord{\left/
 {\vphantom {{j2\pi {d_1}} \lambda }} \right. \kern-\nulldelimiterspace} \lambda })$ correspond to the LOS segments of the AP-to-user link with $d_0$ being the distance between the AP and the user. Additionally, ${{\tilde h}_{rt}} \sim {\mathcal{CN}}(0,1)$ is the NLOS components of the AP-to-user link. Moreover, $K_0$ denotes the Rician factor. Finally, ${\rho _0} = {\rho _r}d_0^{ - {v_0}}$ represent the large-scale path loss of the AP-to-user link. 

\begin{comment}
 Notably, we focus on a general spatially-correlated Rician fading channel. Concretely, we have
 \begin{equation}
{{\bf{H}}_c} = {\bf{R}}_{{\rm{user}}}^{{1 \mathord{\left/
 {\vphantom {1 2}} \right.
 \kern-\nulldelimiterspace} 2}}{\bf{HR}}_{{\rm{AP}}}^{{1 \mathord{\left/
 {\vphantom {1 2}} \right.
 \kern-\nulldelimiterspace} 2}},
 \end{equation}
 \begin{equation}
{{\bf{F}}_c} = {\bf{R}}_{{\rm{RIS}}}^{{1 \mathord{\left/
 {\vphantom {1 2}} \right.
 \kern-\nulldelimiterspace} 2}}{\bf{FR}}_{{\rm{AP}}}^{{1 \mathord{\left/
 {\vphantom {1 2}} \right.
 \kern-\nulldelimiterspace} 2}},
 \end{equation}
 and
  \begin{equation}
{{\bf{G}}_c} = {\bf{R}}_{{\rm{user}}}^{{1 \mathord{\left/
 {\vphantom {1 2}} \right.
 \kern-\nulldelimiterspace} 2}}{\bf{GR}}_{{\rm{RIS}}}^{{1 \mathord{\left/
 {\vphantom {1 2}} \right.
 \kern-\nulldelimiterspace} 2}},
 \end{equation}
 where ${\bf{R}}_{{\rm{AP}}}\in {\mathbb C}^{N_t\times N_t}$, ${\bf{R}}_{{\rm{RIS}}}\in {\mathbb C}^{N\times N}$ and ${\bf{R}}_{{\rm{user}}}\in {\mathbb C}^{N_r\times N_r}$ are the spatially correlation matrices at the AP, active RIS and user, respectively, which are modeled by the commonly used Kronecker correlation model \cite{bjornson2020rayleigh}.
\end{comment}

For transmission, the AP employs the classic V-BLAST technique to transmit a symbol vector ${\bf s}=[s_1,...,s_{N_t}]^T$. Consequently, the transmission rate of the ARIS-ADRM-MIMO is expressed as 
\begin{equation}
R_{\rm{ARIS-ADRM-MIMO}} = N_t\log_2{M}+ \log_2{A}.
\end{equation}

The received signal at the user can be expressed as
\begin{equation}
{\bf{y}} = \sqrt {{P_{{\rm{AP}}}}/{N_t}} ({{\bf{H}}} + {{\bf{G}}}{\bf{\Phi }}{{\bf{F}}}){\bf{s}} + {\bf{n}},
\end{equation}
where ${\bf \Phi}={\rm{diag}}({\boldsymbol{\phi}}_0)={\rm{diag}}([\phi_0(1),...,\phi_0(\tau),...,\phi_0(N)]^T)$ with $\phi_0(\tau)={{\bf a}_k}(l){e^{j{\theta _{\tau}}}}, \tau=1,...,N$. ${\bf n} \in {\mathbb C}^{N_r \times 1}$ is the thermal noise at the user following the distribution ${\cal{CN}}({\bf 0}, {\sigma_0^2}{\bf I}_{N_r})$.

Unlike \eqref{phase_shift}, the phase adjustment of the RIS must take the impact of the direct path into account. Accordingly, the following optimization problem can be formulated as
\begin{equation}\label{phase_optimal}
\begin{array}{*{20}{l}}
{\mathop {\max }\limits_{\bf{\Phi }} }&{{{\left\| {{\bf{H}}} + {{\bf{G}}}{\bf{\Phi }}{{\bf{F}}} \right\|}^2}}\\
{{\rm{s}}{\rm{.t}}{\rm{.}}}&{{\theta _\tau } \in [0,2\pi ),\tau  = 1,...,N}.
\end{array}    
\end{equation}
The optimization problem in \eqref{phase_optimal} can be addressed using various existing techniques, including the semidefinite relaxation (SDR) method \cite{yuan2020intelligent}, the cosine similarity-based approach \cite{yigit2020low}, and the modified block coordinate descent (MBCD) method \cite{kim2022modified}. Therefore, we omit the detailed derivation of obtaining the optimal $\bf \Phi$. Notably, the MBCD method is chosen for simulations due to its ability to deliver comparable performance while significantly reducing computational complexity compared to the SDR method, which is summarized in Algorithm 2, where ${\bf M}_c$ is given by
\begin{equation}\label{MM}
{{\bf{M}}_c} = {\left[ {\begin{array}{*{20}{c}}
{{{\bf{F}}^H}{\rm{diag}}({\bf{g}}_1^H)}&{{\bf{h}}_1^H}\\
 \vdots & \vdots \\
{{{\bf{F}}^H}{\rm{diag}}({\bf{g}}_{{N_r}}^H)}&{{\bf{h}}_{{N_r}}^H}
\end{array}} \right]^H}\left[ {\begin{array}{*{20}{c}}
{{{\bf{F}}^H}{\rm{diag}}({\bf{g}}_1^H)}&{{\bf{h}}_1^H}\\
 \vdots & \vdots \\
{{{\bf{F}}^H}{\rm{diag}}({\bf{g}}_{{N_r}}^H)}&{{\bf{h}}_{{N_r}}^H}
\end{array}} \right],
\end{equation}
where ${\bf{g}}_r$ and  ${\bf{h}}_r$ are the $r$th rows of matrices ${\bf{G}}$ and ${\bf{H}}$, respectively.

At the user, the optimal ML detector can be written as
\begin{equation}
({\bf{\hat s}},\hat k) = \mathop {\arg \min }\limits_{{\bf{s}} \in {\cal S},{{\bf{a}}_k} \in {\cal A}} {\left| {{\bf{y}} - \sqrt {{P_{{\rm{AP}}}}/{N_t}} ({{\bf{H}}} + {{\bf{G}}}{\bf{\Phi }}{{\bf{F}}}}) \right|^2}.
\end{equation}

\begin{algorithm}[t]
\caption{MBCD Method for Obtaining the optimal ${\bf \Phi}$}
\begin{algorithmic}[2]
\REQUIRE ${\bf H}, {\bf G}, {\bf{F}}$ and $I_c$\\
%\ENSURE  $\bf \Phi$
\STATE Initialization: ${\bf{\Psi }}^{(0)} = {\bf 1}_{N+1}$.
\STATE Construct ${\bf{M}}_c$ as \eqref{MM}
\STATE {\textbf{for}} $i = 1:I_c$ {\textbf{do}}
\STATE \quad ${{\bf{\bar \Psi }}^{(i)}} = {{\bf{M}}_c}{{\bf{\Psi }}^{(i - 1)}}$;
\STATE \quad ${\psi ^{(i)}}(n) = {{\bar \psi }^{(i)}}(n)/\left| {{{\bar \psi }^{(i)}}(n)} \right|, n = 1,..,N + 1$;
\STATE \quad ${{\bf{\Psi }}^{(i)}} = {[{\psi ^{(i)}}(1),...,{\psi ^{(i)}}(N + 1)]^T}$.
\STATE {\textbf{end for}}
\STATE ${\boldsymbol{\phi}}'_0 = {\bf{\Psi }}^{(I_c)}/{{\Psi }}^{(I_c)(N+1)}$.
\STATE Output: ${\boldsymbol{\phi}}_0=[{\phi}'_0(1),...,{\phi}'_0(N)]^H$ and ${\bf{\Phi}} = {\rm{diag}}({\boldsymbol{\phi}}_0) $.
\end{algorithmic}
\end{algorithm}

\section{Simulation Results}
In this section, we present comprehensive simulations aimed at evaluating the performance of the proposed ARIS-ADRM scheme and validating our theoretical findings. The simulations are conducted considering both SISO communication systems. The specific simulation parameters are as follows: the noise power is set to $\sigma_r^2=\sigma_0^2=-80$ dBm, the Rician factors for the AP-to-User, AP-to-ARIS and ARIS-to-User links are $K_0 = 0$, $K_1=K_2=3$, respectively. The distance between the AP and user is $d_0=55$ m, the distance between the AP and ARIS is $d_1=5$ m and the distance from the ARIS to user is $d_2=50$ m. The wavelength is $\lambda=0.1$ m, the path-loss exponent of the AP-to-user, the AP-to-ARIS and ARIS-to-User links are $v_0=3.5$, $v_1=v_2=2$, and the path-loss factor is $\rho_r=-30$ dB. The output power of the active RIS is  $P_a = 30$ dBm, and the maximum amplification gain is $\alpha_{\rm{max}}=10$.

\begin{figure}[t]
\centering
\includegraphics[width=0.45\textwidth]{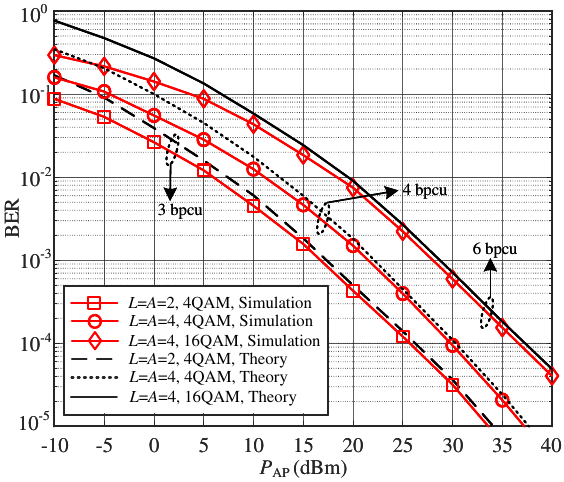}
\caption{Simulation and theoretical results of the proposed ARIS-ADRM scheme under various parameter setting with $N=128$.}\label{BER_Theo}
\end{figure}

\begin{figure}[t]
\centering
\includegraphics[width=0.45\textwidth]{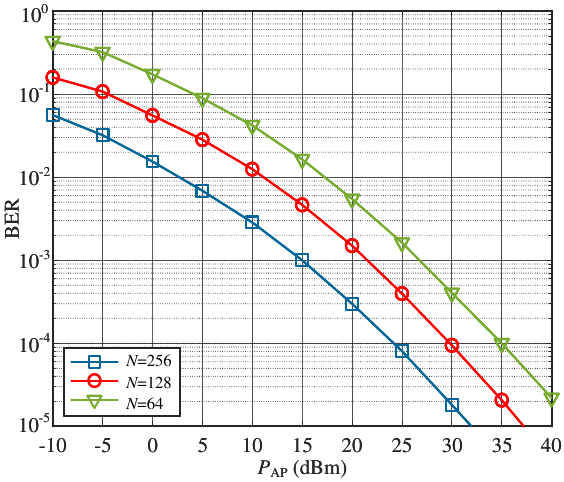}
\caption{BER performance of the proposed ARIS-ADRM scheme with $L=A=4$, 4QAM and various numbers of active RIS REs $N$.}\label{BER_diff_N}
\end{figure}

As illustrated in Fig. \ref{BER_Theo}, the performance of the proposed ARIS-ADRM scheme has been rigorously evaluated across various transmission rate scenarios, encompassing both simulated and analytical average BER results. As shown in Fig. \ref{BER_Theo}, a gap exists between the theoretical and simulated BER results in the low transmit power $P_{\rm{AP}}$ region. This is due to the approximations in the analytical model and the influence of practical factors such as noise dominance and hardware imperfections. As the transmit power $P_{\rm{AP}}$ increases, the analytical results are in perfect alignment with the simulated results. This alignment validates the accuracy and reliability of the analytical models and confirms the robustness of the simulation framework. Furthermore, the theoretical analysis can be served as an effective tool for estimating the BER performance of the proposed scheme and utilized as a critical metric for optimizing system performance.

\begin{figure}[t]
\centering
\includegraphics[width=0.45\textwidth]{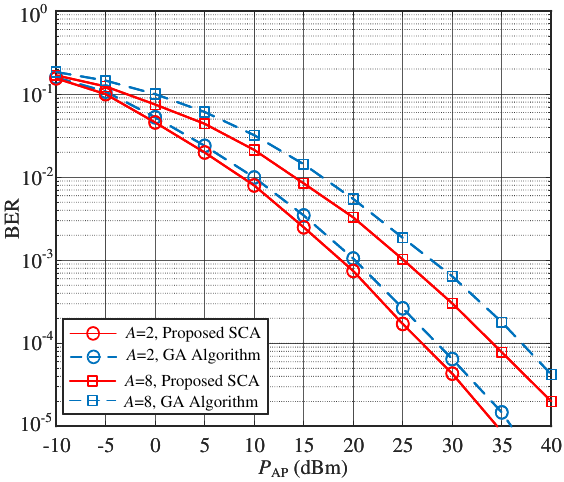}
\caption{BER performance comparison between the proposed SCA-aided AAP codebook design and the GA-adided AAP codebook design schemes with $N=128$, $L=4$, 4QAM and various AAP codebook order $A$.}\label{BER_diff_A}
\end{figure}

Fig. \ref{BER_diff_N} compares the BER performance of the proposed ARIS-ADRM scheme at data rate of 4 bpcu under $L=A=4$, 4QAM and various number of REs $N$. It can be observed from Fig. \ref{BER_diff_N} that the proposed ARIS-ADRM framework with increasing $N$ achieves a significant
enhancement in BER performance. This enhancement is principally attributable to the increased number of active RIS REs, improving the SNR of the received signal. Furthermore, Fig. \ref{BER_diff_A} compares the error performance between the proposed SCA-aided AAP codebook design and the genetic algorithm (GA)-aided AAP codebook design for $N=128$, $L=4$ and 4-QAM with various AAP codebook orders $A$. Our simulations show that the SCA-based AAP codebook outperforms the GA-based one. While GA is effective for global optimization in complex problems, it struggles with the specific structure and constraints of our problem. In contrast, SCA efficiently handles non-convexity, leading to more accurate and computationally efficient solutions.  Additionally, as with conventional multi-level digital modulation techniques, increasing the AAP codebook order results in closer signal levels for the ARIS-ADRM symbols, which degrades the BER performance.  In particular, for a target BER of $10^{-4}$, system with $A=8$ requires approximately 7 dBm higher transmit power compared to the system with $A=2$, highlighting the trade-off between error performance and SE.

\begin{figure}[t]
\centering
\includegraphics[width=0.45\textwidth]{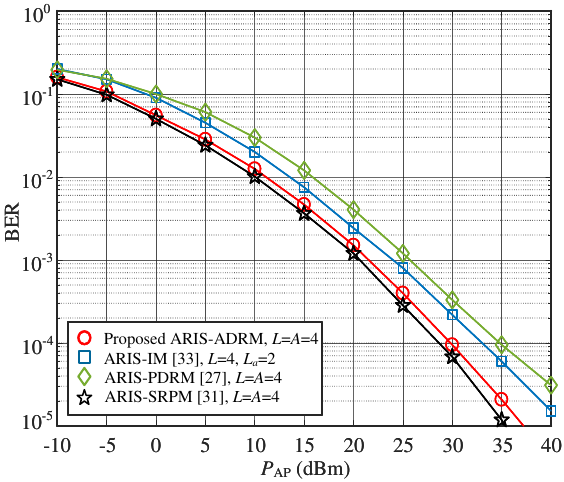}
\caption{BER performance comparisons between the proposed ARIS-ADRM scheme and its benchmark schemes with $N=128$ and 4QAM.}\label{BER_diff_system}
\end{figure}

\begin{figure}[t]
\centering
\includegraphics[width=0.45\textwidth]{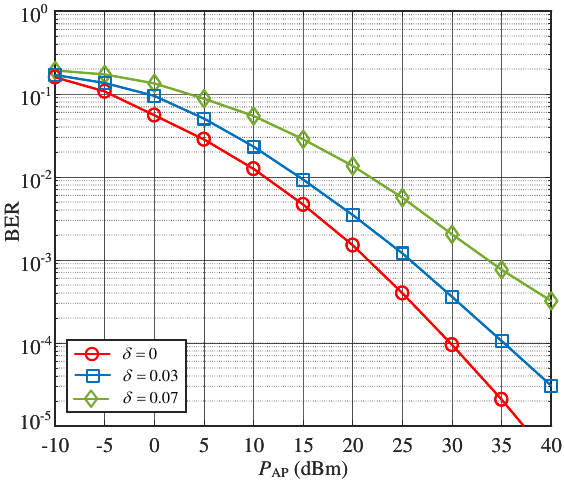}
\caption{BER performance at 4 bpcu ($N=128$, $L=A=4$, and 4QAM) for the proposed ARIS-ADRM scheme in perfect and imperfect CSI cases.}\label{BER_Channel_error}
\end{figure}

\begin{figure}[t]
\centering
\includegraphics[width=0.45\textwidth]{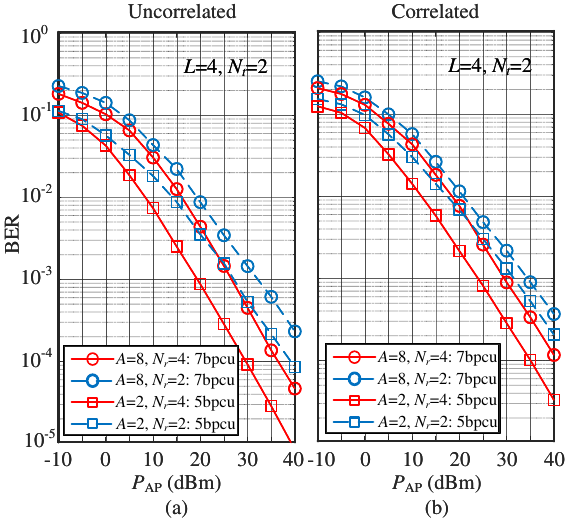}
\caption{Performance of ARIS-ADRM-MIMO with 4QAM under uncorrelated and correlated channel conditions.}\label{MIMO}
\end{figure}

\begin{figure}[t]
\centering
\includegraphics[width=0.45\textwidth]{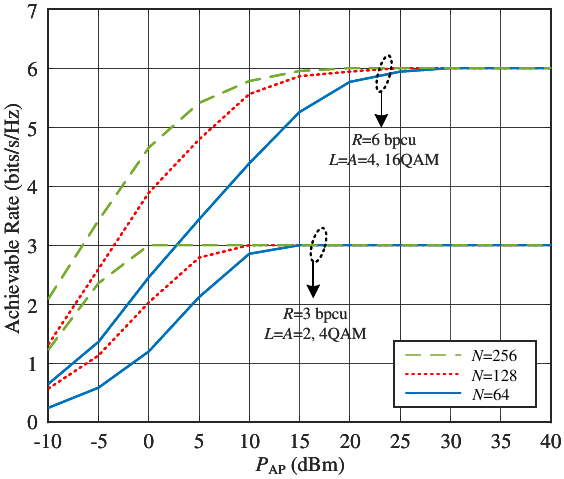}
\caption{Achievable rate of the proposed ARIS-ADRM scheme for different $N$ and $R$.}\label{Achievable_Rate}
\end{figure}

Fig. \ref{BER_diff_system} compares the error performance of the proposed ARIS-ADRM scheme with its benchmark systems under the same transmission rate condition, where the benchmark systems are defined as follows. 
\begin{itemize}
    \item
    ARIS-PDRM \cite{guo2020reflecting}: The active RIS is split into $L$ groups, phase shift patterns are indexed to convey additional information bits. 
    \item
    ARIS-SRPM \cite{yao2023superimposed}: The active RIS is divided into $L$ groups, the superimposed phase offsets are selected from a specific set, according to the extra information to be transferred from the RIS.
    \item
    ARIS-IM \cite{sanila2023joint}: The active RIS is divided into $L$ groups, where $L_a$ out of $L$ groups are switched to ON state to convey index bits. 
\end{itemize}
 Here, we assume the use of an active RIS for the PDRM scheme and SISO configurable for the ARIS-IM scheme to ensure a fair comparison,  despite the fact that passive RISs are predominantly employed in the literature \cite{guo2020reflecting} and MIMO communications are employed in \cite{sanila2023joint}. Moreover, for a fair comparison, the amplification gain of the active REs in the ARIS-PDRM, ARIS-IM and ARIS-SRPM is set to $\alpha^{*} = {\rm min}\{\alpha_{\rm{max}},{\alpha_{\rm{opt}}}\}${\footnote{The same as \eqref{PC_1}, the output power of the active RIS in the ARIS-PDRM/ARIS-IM scheme can be written as ${P_a} = {P_{{\rm{AP}}}}{\alpha ^2}{\left| {{{\bf{p}}^H}} \right|^2} + \sigma _r^2{\alpha ^2}$. Thus, the optimal amplification factor can be obtained as ${\alpha_{\rm{opt}} } = \sqrt {\frac{{{P_a}}}{{{P_{{\rm{AP}}}}{{\left| {{{\bf{p}}^H}} \right|}^2} + \sigma _r^2}}}$.}} \cite{sanila2023joint}. In the case of $N=128$, $L=4$, and 4QAM, we set the simulation parameters as $A=4$ in the proposed ARIS-ADRM and ARIS-SRPM schemes, $L_a=2$ in the ARIS-PDRM and ARIS-IM schemes, to achieve the transmission rate of 4 bpcu. It can be observed from Fig. \ref{BER_diff_system} that the proposed ARIS-ADRM scheme significantly outperforms the conventional ARIS-IM and ARIS-PDRM schemes in the whole transmit power region. In particular, for a target BER of $10^{-4}$, the proposed scheme offers 3 dB and 5 dB transmit power gains compared to the ARIS-IM and ARIS-PDRM schemes, respectively. This is because the proposed ARIS-ADRM scheme is able to fully exploit the additional amplitude domain DoF provided by the active RIS while maintaining the benefits of IM. Although the proposed system experiences slight performance loss compared to the ARIS-SRPM system, it requires less overall transmission energy. This is because, in the ARIS-SRPM system, all REs are activated, and the amplitude of each RE is set to its maximum value, resulting in the highest overall transmission energy.

Fig. \ref{BER_Channel_error} evaluates the impact of channel estimation errors on the BER performance of the proposed ARIS-ADRM scheme. Here, the imperfect channel vectors are modeled as
${\bf f}_e = {\bf f}+{\bf f}_{\rm t}$ and ${\bf g}_e = {\bf g}+{\bf g}_{\rm r}$, where ${\bf f}_{\rm t}$ and ${\bf g}_{\rm r}$ are the channel error with its entries following ${\cal {CN}}(0,{\delta}^2)$. It is worth mentioning that the performance of the perfect CSI case (i.e., $\delta=0$) is also included to facilitate comparative analysis. It can be seen from Fig. \ref{BER_Channel_error} that the error performance of the proposed ARIS-ADRM scheme deteriorates with an increase in $\delta$. Specifically, when BER is $10^{-3}$, the transmit power deterioration of the proposed scheme is 4 dB and 12 dB for $\delta=0.03$ and $\delta=0.07$, respectively, compared to the perfect CSI case. This demonstrates the sensitivity of the proposed scheme to
CSI errors and highlights the importance of accurate channel estimation.

Fig. \ref{MIMO} depicts the performance of the proposed ARIS-ADRM-MIMO scheme under various parameter configurations with 4QAM. The parameters are set as $L=4, A=8, N_t=2$ and $L=4, A=2, N_t=2$, corresponding to transmission rates of 7 bpcu and 5 bpcu, respectively. It can be seen from Fig. \ref{MIMO} (a) that for both 5 bpcu and 7 bpcu cases, increasing the number of receive antennas improves the diversity order and the BER performance. This improvement is attributed to the increased spatial diversity provided by additional receive antennas, which mitigates the effects of fading and enhances signal detection accuracy. To reflect more realistic conditions, spatial correlations are considered in Fig. \ref{MIMO} (b), modeled using the widely recognized Kronecker model. The correlated Rician channels of each link are represented as ${{\bf{H}}_c} = {\bf{\bar H}} + {\bf{\tilde H}} = {\bf{\bar H}} + {\bf{R}}_{{\rm{user}}}^{\frac{1}{2}}{\bf{\tilde HR}}_{{\rm{AP}}}^{\frac{1}{2}}$, ${{\bf{F}}_c} =  {\bf{\bar F}} + {\bf{R}}_{{\rm{RIS}}}^{\frac{1}{2}}{\bf{\tilde FR}}_{{\rm{AP}}}^{\frac{1}{2}}$ and ${{\bf{G}}_c}  = {\bf{\bar G}} + {\bf{R}}_{{\rm{user}}}^{\frac{1}{2}}{\bf{\tilde GR}}_{{\rm{RIS}}}^{\frac{1}{2}}$, respectively, where ${\bf{R}}_{{\rm{AP}}}\in {\mathbb C}^{N_t\times N_t}$, ${\bf{R}}_{{\rm{RIS}}}\in {\mathbb C}^{N\times N}$ and ${\bf{R}}_{{\rm{user}}}\in {\mathbb C}^{N_r\times N_r}$ are the spatially correlation matrices at the AP, active RIS and user, respectively, which are modeled by the commonly used Kronecker correlation model \cite{bjornson2020rayleigh}.The same trend is observed in the correlated channel conditions, further validating the efficiency of our system under varying channel environments.

Fig. \ref{Achievable_Rate} presents the achievable rate of the proposed ARID-ADRM scheme under various $N$ and $R$ scenarios. In this context, the active RIS with $N=64,128$, and 256 REs are divided into $L=2$, and 4 groups, achieving transmission rates of $R=3$ and 6 bpcu, respectively. The results indicate that increasing the number of REs, $N$, is beneficial for more rapidly converging to the target data rate. This is because the additional REs can enhance signal power, thereby increasing the transmission rate. Consequently, a larger $N$ not only accelerates the rate convergence but also enhances overall system performance by fully utilizing available resources.

\section{Conclusion}
In this paper, we proposed a novel ARIS-ADRM transmission scheme, which leverages the benefits of active RIS and IM to enhance SE without incurring additional RF chain costs. Our comprehensive theoretical analysis included a closed-form expression for the upper bound on the ABEP and an achievable rate analysis, providing valuable insights into the performance limits and capabilities of the proposed scheme. Furthermore, we formulated an optimization problem for constructing the AAP codebook to minimize ABEP, applying the SCA method to handle the non-convex QCQP efficiently. Simulation results confirmed that the ARIS-ADRM scheme significantly improves error performance under the same SE conditions compared to its benchmark schemes, due to its flexible adaptation of the transmission rate by fully exploiting amplitude domain DoF provided by active RIS. These findings substantiate the potential of ARIS-ADRM in advancing communication systems by optimizing resource utilization and enhancing overall system performance.

\begin{appendices}
\section{The calculation of noise power in (8)} 
Let us first review the basics of probability theory. Assume that  $X$ and $Y$ are two independent random variables, the mean and variance of the product $Z=XY$ are calculated as
\begin{equation}
\left\{ {\begin{array}{*{20}{l}}
{{\mu _Z} = {\mu _X}{\mu _Y},}\\
{\sigma _Z^2 = \sigma _X^2\sigma _Y^2 + \mu _X^2\sigma _Y^2 + \mu _Y^2\sigma _X^2}.
\end{array}} \right.
\end{equation}
In addition, the mean and variance of the sum $W=Z+Y$ are given by
\begin{equation}
\left\{ {\begin{array}{*{20}{l}}
{\mu _W} = {\mu _X}+{\mu _Y},\\
\sigma _W^2 = \sigma _X^2 + \sigma _Y^2.
\end{array}} \right.
\end{equation}

Let $\tilde n = \sum\nolimits_{l = 1}^L {\sum\nolimits_{i = 1}^{\bar N} {{\bf{a}}_k(l)\left| {{g_{li}}} \right|{e^{ - j{\phi _{li}}}}{n_r}} }$ for ease of notation, the equivalent noise $w$ can be thus simplified as $w = \tilde n + n$. According to the above theory, the mean and variance of $w$ can be calculated by
\begin{equation}
 {\mu _w} = {\mathbb E}\left[ {\tilde n + n} \right]={\mathbb E}\left[ {\tilde n }\right]+{\mathbb E}\left[ {n }\right] = 0,
\end{equation}
and
\begin{equation}
\begin{aligned}
\sigma _w^2 &= {\mathbb E}\left[ {{{(w - {\mu _w})}^2}} \right] \\
&= {\mathbb E}\left[ {{{\tilde n}^2} + {n^2} + 2\Re \{ \tilde nn\} } \right]\\
& = {\mathbb E}\left[ {{{\tilde n}^2}} \right] + {\mathbb E}\left[ {{n^2}} \right]\\
& = N{\rho _2}\sigma _r^2{{\left( {\sum\nolimits_{l = 1}^L {{{({{\bf{a}}_k}(l))}^2}} } \right)} \mathord{\left/
 {\vphantom {{\left( {\sum\nolimits_{l = 1}^L {{{({{\bf{a}}_k}(l))}^2}} } \right)} L}} \right.
 \kern-\nulldelimiterspace} L} + \sigma _0^2.
\end{aligned}
\end{equation}

\section{The calculation of the mean  and variance of for random variable $h_l$}
The channel magnitudes of $\left| {{f_{li}}} \right|$ and $\left| {{g_{li}}} \right|$ are independent Rician distributed variables with the means and variances given by 
\begin{equation}
\left\{ {\begin{array}{*{20}{l}}
{{\mu _{\left| {{f_{li}}} \right|}} = \frac{1}{2}\sqrt {\frac{{{\rho _1}\pi }}{{1 + {K_1}}}} {e^{ - \frac{{{K_1}}}{2}}}\left[ {\left( {1 + {K_1}} \right){I_0}\left( {\frac{{{K_1}}}{2}} \right) + {K_1}{I_1}\left( {\frac{{{K_1}}}{2}} \right)} \right]},\\
{\sigma _{\left| {{f_{li}}} \right|}^2 = {\rho _1} - \mu _{\left| {{f_{li}}} \right|}^2},
\end{array}} \right.
\end{equation}
and
\begin{equation}
\left\{ {\begin{array}{*{20}{l}}
{{\mu _{\left| {{g_{li}}} \right|}} = \frac{1}{2}\sqrt {\frac{{{\rho _2}\pi }}{{1 + {K_2}}}} {e^{ - \frac{{{K_2}}}{2}}}\left[ {\left( {1 + {K_2}} \right){I_0}\left( {\frac{{{K_2}}}{2}} \right) + {K_2}{I_1}\left( {\frac{{{K_2}}}{2}} \right)} \right]},\\
{\sigma _{\left| {{g_{li}}} \right|}^2 = {\rho _2} - \mu _{\left| {{g_{li}}} \right|}^2},
\end{array}} \right.
\end{equation}
respectively, where $I_0(\cdot)$ and $I_1(\cdot)$ are the modified Bessel functions of the first kind  of order 0 and order 1, respectively.

According to CLT, as the number of active RIS REs $N$ increases, $h_l=\sum\nolimits_{i = 1}^{\bar N} {\left| {{f_{li}}} \right|\left| {{g_{li}}} \right|}$ can be treated as a Gaussian random variable with mean $\mu$ and variance $\sigma^2$, where
\begin{equation}\label{mean_1}
\mu  = \bar N{\mu _{\left| {{f_{li}}} \right|}}{\mu _{\left| {{g_{li}}} \right|}},
\end{equation}
\begin{equation}\label{variance_1}
{\sigma ^2} = \bar N({\mu _{{{\left| {{f_{li}}} \right|}^2}}}{\mu _{{{\left| {{g_{li}}} \right|}^2}}} - \mu _{\left| {{f_{li}}} \right|}^2\mu _{\left| {{g_{li}}} \right|}^2).
\end{equation}
Since ${\mu _{{{\left| {{f_{li}}} \right|}^2}}} = \sigma _{\left| {{f_{li}}} \right|}^2 + \mu _{\left| {{f_{li}}} \right|}^2 = {\rho _1}$ and ${\mu _{{{\left| {{g_{li}}} \right|}^2}}} = \sigma _{\left| {{g_{li}}} \right|}^2 + \mu _{\left| {{g_{li}}} \right|}^2 = {\rho _2}$, the variance $\sigma^2$ in \eqref{variance_1} can be further calculated as
\begin{equation}
{\sigma ^2} = \bar N({\rho _1}{\rho _2} - \mu _{\left| {{f_{li}}} \right|}^2\mu _{\left| {{g_{li}}} \right|}^2).  
\end{equation}

\end{appendices}

\bibliography{ARIS_ADRM}
\bibliographystyle{ieeetr}

\end{document}